%% file: dns-arxive.tex
\documentclass[10pt]{article}  
\usepackage{cite}
\usepackage{amsmath,amssymb,amsfonts}
\usepackage{algorithmic}
\usepackage{graphicx}
\usepackage{textcomp}
\usepackage{xcolor}
\def\BibTeX{{\rm B\kern-.05em{\sc i\kern-.025em b}\kern-.08em
    T\kern-.1667em\lower.7ex\hbox{E}\kern-.125emX}}

\usepackage{tikz}
\usepackage{paralist}
\usepackage{wrapfig}
\usepackage{xspace}
\usepackage{url}
\usepackage{xurl}
\usepackage{color}
\usepackage{comment}
\usepackage{enumitem}
\usetikzlibrary{graphs,graphs.standard,shapes,calc}
\usepackage{soul}
\usepackage{algorithm}
\usepackage{algorithmic}
\usepackage{stfloats}
\usepackage{accents} %%%%%%%%%%%%%%%%%%%%%
\newcommand{\ariel}[1]{{\color{blue} #1}\normalcolor}

\newcommand{\delete}[1]{{\color{green} #1}\normalcolor}
\newcommand\ignore[1]{}
\newcommand\nsdionly[1]{}
\newcommand{\DNSUD}{DNS{\em {RU}}}
\newcommand{\UpdateDB}{UpdateDB}
\newcommand{\Extension}{DNSRU-ext}
\newcommand{\UpdateDBext}{UpdateDB-ext}
\newcommand{\dynamic}{\emph{dynamic}}
\begin{document}
%\title{\DNSUD{}: Real-Time DNS Update}
\title{Decoupling DNS Update Timing from TTL Values}

\ignore{%%%%%%%%%%%%%%%%
%authors
\author{\IEEEauthorblockN{\rm Author1}\\
Institution1
\and
{\rm Author2}\\
Institution2
}
}%%%%%%%%%%%%%%%%

\author{
{Yehuda Afek}\\
{\textit{Tel-Aviv University}}
\and
{Ariel Litmanovich}\\
{\textit{Tel-Aviv University}}
}

\date{\ }
\maketitle

\pagenumbering{arabic}
\input{abstract}

%%%\begin{IEEEkeywords}
%%%component, formatting, style, styling, insert
%%%\end{IEEEkeywords}

%-------------------------------------------------------------------------------
\input {intro}
\input {background}
\input {solution}
\input {measurements}
\input {analysis}
\input {security}

\input {drill-down}
\input {client}
\input {discussion}
\input {conclusion}
\input {acknowledgements}
%-------------------------------------------------------------------------------
%%%%\input{dns2.bib}
\bibliographystyle{plainurl} % We choose the "plain" reference style
\bibliography{dns-arxive} % Entries are in the refs.bib file
%%%%\vspace{12pt}

\end{document}

%% file: abstract.tex
\begin{abstract}
\ignore{%%%%%%%%%%%%%%%
We propose DNS Real-time Update (\DNSUD{}), a new DNS service that enables secure, real-time updates of cached domain records in resolvers around the glob, even before the associated Time To Live (TTL) values expire.
}%%%%%%%%%%%%%%%
A relatively simple safety-belt mechanism for improving DNS system availability and efficiency is proposed here.
While it may seem ambitious, a careful examination shows it is both feasible and beneficial for the DNS system.
The mechanism called `DNS Real-time Update' (\DNSUD{}), a service that facilitates real-time and secure updates of cached domain records in DNS resolvers worldwide, even before the expiration of the corresponding Time To Live (TTL) values.

\ignore{%%%%%%%%
This service allows Internet domain owners to update any of their domain records in the resolvers' cache worldwide whenever necessary, regardless of the TTL value assigned to the domain.
}%%%%%%%%
This service allows Internet domain owners to quickly rectify any erroneous global IP address distribution, even if a long TTL value is associated with it.
By addressing this critical DNS high availability issue, \DNSUD{} eliminates the need for short TTL values and their associated drawbacks. 
Therefore, \DNSUD{} DNSRU reduces the traffic load on authoritative servers while enhancing the system's fault tolerance.
\ignore{%%%
Additionally, we complement the service and propose methods to pull domain record updates from resolvers to client caches before their TTL expires on the client side.
}%%%
In this paper we show that our \DNSUD{} design is backward compatible, supports gradual deployment, secure, efficient, and feasible.

%%%%%%%%%%%%%%%%%%%%%%%%%%%%%%%%%%%%%%%%%%%%%%%%%%%%%

\ignore{\DNSUD{} (DNS Update), a new DNS service is proposed here to enable secure real-time updates of cached domain records in resolvers by authoritative serves, before the corresponding TTLs have expired. 
Thus enabling Internet services to update their domain records in resolver caches in case of a record with the wrong IP address and with a large TTL has been mistakenly distributed. 
This eliminates the main motivation for short TTL values, allowing larger TTL values that reduce the traffic load on authoritative servers and the latency experienced by clients.}%%%%%%%%%

\ignore{%%%%%
\DNSUD{} (DNS Update), a new DNS service is proposed here to enjoy the benefits of large TTLs without their disadvantages in the DNS system.
\DNSUD{} enables secure real-time updates from authoritative to resolver servers about IP address changes that have occurred before the TTL of the corresponding domains expires.
Thus providing domain owners a way to update their records in cases the wrong IP has been (mistakenly) distributed with a large TTL. Furthermore, by allowing large TTL values without their disadvantages, the traffic load on the authoritative servers and the latency experienced by clients is reduced.
}%%%%%%%%%%

\ignore{ %% not abstract material should go to introduction
DNS extensively uses the cache mechanism to improve performance. The control on the caching mechanism is done through \textit{Time-to-Live}(TTL) values. 
On the one hand, there is a dual motivation to increase the TTL values: to reduce the load on the authoritative servers and the latency experienced by clients. On the other hand, if there is a problem with the DNS service or change of the website's IP address, the availability of that website is compromised until the TTL has expired. This problem is not theoretical and even occurred in reality. }%%%

\ignore{ %%%%%%%%%OLD ABSTRACT 
This paper suggests adding a service called \textbf{LITZMAN} - Local IP Transfer Zone Manager to the DNS system.
This service allows real-time notifications from authoritative to Resolver servers about IP changes to some domains.
This solution is not only a lifeline to the DNS network, but also allows to reduce the required traffic between Resolvers and authoritative servers by increasing the TTL values and updating our service in case of changes.
We will discuss in this paper the various considerations that led to the realization of the solution and its extensions.
DNS extensively uses the cache mechanism to improve performance. The control on the caching mechanism is done through \textit{Time-to-Live}(TTL) values. 
On the one hand, there is a dual motivation to increase the TTL values: to reduce the load on the authoritative servers and the latency experienced by clients. On the other hand, if there is a problem with the DNS service or change of the website's IP address, the availability of that website is compromised until the TTL has expired. This problem is not theoretical and even occurred in reality. 
}%%%%%%%%% 

\end{abstract}

%%% Local Variables:
%%% mode: latex
%%% TeX-master: "cwc"
%%% End:

%% file: intro.tex
\section{Introduction}
\label{sec:intro}

A somewhat simple and radical safety belt mechanism for the DNS system high availability is proposed here.
While the suggestion seems ambitious and bold, we carefully examine all its aspects and argue its feasibility and advantages for the DNS system. 

High availability is crucial for the DNS system since its unavailability renders the Internet practically useless, making it impossible to access corresponding services.
To ensure the high availability of their domains, domain owners usually associate them with very low TTL values \cite{choose-ttl-values}.
However, low TTL values can be a double-edged sword. While they might impair availability during a DDoS attack on the authoritative server, they can also increase traffic load and reduce the effectiveness of the DNS cache.
Consequently, this leads to higher operational costs for the system.
\ignore{
potentially impairing availability while also increasing traffic load and reducing the effectiveness of the DNS cache}
Consequently, this leads to higher operational costs for the system.

%%%%%%%%%%%

\subsection{Motivation}

Consider the following two scenarios at the highly profitable Internet Foo-bar shop, which generates millions of dollars in daily revenue from the sales of foo-bars.
The first scenario took place on October 10, 2016, when the domain admin of foo-bar.com mistakenly updated the domain with an incorrect IP address and associated it with a TTL value of twelve hours.
As a result, there was a rapid decline in traffic and sales on the foo-bar on-line shop.
Despite the admin's urgent efforts to rectify the mistake by updating the authoritative servers with the correct record, the problem persisted.
Many resolvers worldwide continued to retain the erroneous record throughout the duration of the twelve-hour TTL.
Consequently, the admin found herself helpless, ultimately facing reprimand and termination by her boss.

The second scenario happened just a few days later when a new admin was hired to manage foo-bar.com and implemented a strict low TTL value of just 2 minutes.
This ensured that any potential loss in business would not exceed a mere 2 minutes.
However, on October 21, 2016, foo-bar.com fell victim to the Mirai DDoS attack, which affected the Dyn ISP that hosts the foo-bar authoritative servers.
Within 2 to 3 minutes of the attack, the traffic and sales on foo-bar servers drastically diminished as resolvers worldwide promptly cleared the foo-bar.com record from their caches once the $2$-minute TTL values expired, and could not retrieve a new update from the DDoSed authoritative server.
\ignore{Reprimand and termination was his fate as well.}
His fate was the same as his predecessor: reprimand and termination.
While alternative patches may be suggested, such as using stale records when the authoritative is unreachable, each patch is likely to prolong the back-and-forth struggle, potentially leading to more devastating scenarios.

\ignore{%%%%%%%%%%%%%%%%%%%%%
Unfortunately both scenarios occasionally happen.
For example, the first scenario occurred in 2019 resulting in Microsoft outage in which Azure and other services were unavailable for a few hours due to an erroneous DNS update that could not be promptly corrected or undone \cite{azure}.
Other similar incidents have been reported, e.g., \cite{siteground,zoho,slack,win-update,akami}.
While the second scenario has occurred uring the 2016 Dyn DDoS attack, when the authoritative DNS servers went down, domains with short TTL values were affected almost instantly \cite{dyn}.
A similar scenario has happened with Facebook \cite{facebook}. Moreover, short TTL values are expensive \cite{moura2019cache} in terms of futile DNS traffic generated by the constant refreshing of popular domain records in various resolvers every TTL time. 
}%%%%%%%%%%%%%%%%%%%%%
\subsection{Background}
Unfortunately, both of these scenarios have occurred in practice.
For instance, in 2019, a scenario similar to the first occurred, resulting in an outage of several Microsoft services.
Azure and other services were inaccessible for several hours due to an erroneous DNS update that couldn't be promptly rectified or undone \cite{azure}.
Similar incidents have been reported with other companies such as SiteGround, Zoho, Slack, Windows Update, and Akamai \cite{siteground,zoho,slack,win-update,akami}. 
We believe that many more such incidents go unreported.
An instance of the second scenario transpired during the 2016 Mirai DDoS attack, which affected Dyn and exposed the vulnerability of domains with short TTL values. When the authoritative DNS servers hosted by Dyn went down, domains with low TTL values such as Twitter, Netflix, the Guardian, Reddit, and CNN experienced almost instantaneous impact \cite{dyn}. A comparable situation unfolded with Facebook, as reported in \cite{facebook}.
Additionally, maintaining short TTL values is costly \cite{moura2019cache}, as it leads to unnecessary DNS traffic generated by frequent refreshing of popular domain records in various resolvers, which occurs every TTL duration.

\ignore{%%%%%%%%%%%
These issues arise from the inherent nature of the DNS system, which is a "Pull" rather than a "Push" system.
In this system, clients and resolvers actively retrieve information from authoritative servers instead of the authoritative servers proactively pushing the data to them.
The proposed \DNSUD{} service introduces a push element into this fundamental pull-based system, thereby addressing the limitations associated with the traditional "Pull" approach.
}%%%%%%%%%%%

%%%%%%%%%%%%%%%%%%%%%

\subsection{\DNSUD{}: resolution for concerns} 

In this paper, we introduce \DNSUD{} a method that combines the advantages of an undo command for quick recovery from erroneous updates with the benefits of large TTL values. 
While \DNSUD{} changes the DNS system, it is backward compatible supportes gradual deployment and comes with several advantages.
\DNSUD{} facilitates the timely update of records in resolver caches worldwide, even before the associated TTL expires.
By notifying resolvers about recently updated domains, \DNSUD{} ensures that the corresponding records are promptly removed from their cache.
If a removed domain is queried, the resolver performs a new resolution in the standard DNS system to obtain the latest update.

This paper examines the following aspects, demonstrating the feasibility of \DNSUD{} (for a full version see \cite{arielThesis})
%%In this paper we address each of the following properties in seperate sections:
%%Our design of \DNSUD{} possesses the following properties:
%%\begin{enumerate}[leftmargin=0.12in]\setlength\itemsep{0em}
%%%\begin{itemize}[leftmargin=0.12in]\setlength\itemsep{0.25em}
    %%\item[\textbf{ Enable large TTL values:}]
    \begin{itemize}
        \item 
    \textbf{Enable large TTL values:} 
    \DNSUD{} allows Internet services to significantly increase the TTL values associated with their domain names in resolver caches (client caches are discussed in the sequel).
    To allow the association of a much larger TTL value with domain names, we insert a new TTL field, in addition to the traditional one, called \DNSUD-TTL (Section \ref{sec:increase-ttl}).
    \ignore{In this way, we are able to save a lot of futile traffic between resolvers and authoritative servers.}
    This in turn may eliminate most of the futile traffic between resolvers and authoritative servers.
    In Section \ref{sec:beyond}, we expand \DNSUD{} to enable clients to utilize the additional larger TTLs, thereby reducing futile DNS traffic also between clients and resolvers.

    \item
    \textbf{Backward compatibility and gradual deployment:} 
    To ensure backward compatibility, the \DNSUD-TTL is included in the ADDITIONAL optional section \cite{rfc1035} of DNS responses.
    This way, an authoritative server that uses \DNSUD{} can communicate with resolvers that have implemented \DNSUD, as well as with those that have not. Similarly, a resolver that uses \DNSUD{} can communicate with both authoritative servers that have implemented \DNSUD{} and those that have not (Section \ref{sec:increase-ttl}).
    Additionally, \DNSUD{} supports DNS routing features including the DNS unicast, anycast routing and round-robin DNS \cite{unicast, roundrobin}.

    \item
    \textbf{Feasible:}
    We show that \DNSUD{} can be efficiently useful for many domains (hundreds of millions), including Content Delivery Networks (CDNs) \cite{centralized-dns}. A key idea behind \DNSUD{} is that a lot of domains on the Internet do not change their IP addresses very often.
    We observed that around 80\% of domains change their IP addresses less frequently than once a month on average (shown in Figure \ref{fig:domains-update}). We call these \emph{stable} domains. 
    In the \DNSUD{} design, our primary focus is on these \emph{stable} domains; other domains are presumed to persist with the standard DNS system.
    It is noteworthy that many domains supported by big CDNs are also \emph{stable}. This trend is growing as CDNs like Cloudflare and Fastly transition to fewer but larger data centers, leveraging anycast routing amongst them, instead of frequently updating domains IP addresses as in traditional CDN networks.
    Moreover, many \emph{stable} domains possess short TTL values, which, in fact, amplifies the advantages of \DNSUD, making it even more beneficial.
    
    \item
    \textbf{Efficient and scalable:} 
    \DNSUD{} can efficiently serve hundreds of millions of domains, including an efficient update of subdomains within a specific domain name.
    Furthermore, because \DNSUD{} services \emph{stable} domains, the load on our system and the storage required for the updated domains are moderate, eliminating the need for complex scalability mechanisms (Section \ref{sec:solution-usage}).
    
    \item
    \textbf{Secure:} 
    \DNSUD{} does not compromise the security of the DNS system. We take measures to prevent \DNSUD{} from becoming a new single point of failure. A detailed security analysis, along with mitigations for potential attacks on our system, is presented in Section \ref{sec:security}.
\end{itemize}

In addition to the above properties, the following considerations make \DNSUD{} attractive to authoritatives and resolvers:
%%%\begin{itemize}[leftmargin=0.12in]\setlength\itemsep{0.25em}
\begin{itemize}
    \item 
    \textbf{Timely:}
    \ignore{%%%%%%
    There reasons why it is the right time for \DNSUD{}. Firstly, the push for high availability. Secondly, the large percentage of stable domains in general, and because big CDNs tending to use fewer and larger data centers with smarter anycast routing among them, instead of changing IP addresses.
    
    There are three key reasons why it is the right time for \DNSUD{}. Firstly, the push for high availability, as discussed earlier. Secondly, the large percentage of stable domains in general, and because big CDNs are using fewer and larger data centers with smarter anycast routing among them, instead of changing IP addresses.
  
    Three factors make the time ready for \DNSUD{}; firstly, the high-availability motivation as explained earlier, secondly, the high percentage of stable domains in general, and the trend of large CDNs to move to fewer and larger data-centers using anycast routing rather than changing IP addresses.
    }%%%%%%
    The \DNSUD{} is a timely development in the DNS system, not only because it solves a high-availability issue, and increases the DNS efficiency, but also due to the high and increasing percentage of stable domains in the Internet. 
    The percentage of \emph{stable} domains (those that change their IP addresses less frequently than once a month on average) is above $80\%$ and keeps growing (see Section \ref{sec:measurements}). 
    Thus making \DNSUD{} relevant to over $80\%$ of the domains in the Internet.
    This trend can be attributed among others to the shift by major CDNs towards consolidating into fewer but larger data-centers that utilize anycast routing among them (e.g., CloudFlare and Fastly), instead of relying on thousands of reverse proxies with distinct IP addresses \cite{cdn-market}.
    Additionally, the market share of CDNs increased by more than $400\%$ between the years $2016$ and $2023$ \cite{cdn-market}, particularly for those CDNs (Cloudflare and Fastly) that a majority of their domains are \emph{stable} domains.
    
    \item
    \textbf{Economic implications and incentives:}
    \ignore{
    The introduction of \DNSUD{} has economic implications for both authoritative servers and resolvers.
    This approach can eliminate revenue losses in the current DNS system and even create new revenue streams for authoritative servers, see Section \ref{sec:economic-implications}.
    \ariel{
    To ensure effective management, we recommend that the \DNSUD{} servers will be handled in a manner similar to the root servers.}
    }
    Introducing \DNSUD{} offers financial benefits for both resolvers and authoritative  servers operators.
    It can reduce revenue loss and create new income opportunities (see Section \ref{sec:economic-implications}). 
    We recommend integrating the \DNSUD{} service into the root DNS servers.
    %%%%We recommend managing \DNSUD{} servers in the same manner as root servers.
\end{itemize}

%%%%%%%%%%%%%%%%%%%%%%%%%

\nsdionly{%%%%%%%%%%%%%%%%%%%%%%%%%%%
After presenting the \DNSUD{} architecture, this paper addresses the following additional challenges:
\begin{itemize}[leftmargin=0.12in]\setlength\itemsep{0.25em}
    \item 
    We discuss the modifications necessary in resolvers to ensure backward compatibility and gradual adoption of \DNSUD{}.
    \ignore{This allows different parts of the DNS system to use \DNSUD{} while maintaining traditional operation in other parts.}
    \item
    We investigate the applicability of \DNSUD{} to a wide range of domains, including Content Delivery Networks (CDNs) \cite{kazunori2012cdn, centralized-dns}.
    \item
    We assess the feasibility of \DNSUD{} to serve hundreds of millions of domains efficiently.
    \item
    We explore strategies to eliminate futile traffic and improve overall system efficiency, given \DNSUD{}.
    \item
    We address the efficient update of subdomains within a specific domain name.
    \item
    We analyze and ensure that \DNSUD{} does not compromise the security of the DNS system. 
    \item
    We take measures to prevent \DNSUD{} from becoming a new single point of failure in the DNS system.
    \item
    We evaluate the economic costs and benefits of \DNSUD{}.
    \item
    We explore methods to incentive resolvers to implement \DNSUD{}.
\end{itemize}
}%%%%%%%%%%%%%%%%%%%%%%%%%%%%%%%%%%%%%%%%%%%%%%%%%%%%%%%%%%%%%%%

\ignore{%%%%%%%
Examining the relevance of the \DNSUD{} method provides a primary observation about the severity of the current futile DNS traffic.
}%%%%%%%

\ignore{%%%%%%%%%%%%%%%%%%%%%%%%%%%%%%%%%%%%%%
According to our measurements, about $80\%$ of the domains on the Internet change their IP address infrequently, less than once a month on average. Let us call such domains \emph{stable} domains.

Let us define domain update time, $T^D_{update}$ as the average inter update time for domain $D$. We say domain $D$ is \emph{stable} if $T^D_{update} > 1$ month.

\DNSUD{} is designed under the assumption that only \emph{stable} domains use the service, while the others keep using the traditional system.
Moreover, most of them have a small TTL value.
The observation is that \emph{stable} domains have a very high ratio of domain update time to TTL length. As a typical example, consider the popular Microsoft \emph{stable} domain office.com whose TTL is $5$ minutes, but its IP address has not been updated for $3$ years.
Thus, at least every $5$ minutes, office.com is evicted from the resolvers' cache, and upon receiving a DNS request, the resolvers issue a new query to the authoritative server holding office.com.
Surprisingly, most domains serviced by Cloudflare and Fastly, two of the largest CDNs, are \emph{stable}.
}%%%%%%%%%%%%%%%%%%%%%%%%%%%%%%%%%%%%%%%%%%%%%%%%%%%

%%%%%%%%%%%%%%%%%%%%%%%%%%%%

\ignore{While the method could be applicable also for clients and stub resolvers, in their case efficiency and complexity of the method, as well as their small cache size, needs to be considered, see Section \ref{sec:beyond}.}
\nsdionly{%%%%%%%%%%%%%%%%%%
Unlike the phone book, the trigger in \DNSUD{} is initiated by the authoritative server rather than the client, i.e., some kind of a pull method initiated by a push.
}%%%%%%%%%%%%%%%%%%

Related work and alternative methods are presented in Section \ref{sec:background}. 
The \DNSUD{} service is described in Section \ref{sec:solution}.
In section \ref{sec:measurements}, we provide measurements on domain IP address update times and their TTL values.   
In section \ref{sec:analysis}, we analyze the relevancy and feasibility of \DNSUD. 
In section \ref{sec:security}, we present the security threat analysis and the corresponding mitigations. 
%%%In section \ref{sec:security} we discuss security mechanisms 
%%%required to mitigate any security threat introduced by the method.
\ignore{
A detailed description of the \DNSUD{} is provided in Appendix \ref{sec:drill-down}.
}
%%%%%QQQQ Appendix to Section:  A detailed description of the \DNSUD{} is provided in Section \ref{sec:drill-down}.
In Section \ref{sec:beyond}, we discuss methods for client browsers to pull new domain updates from resolvers before their TTL has expired.
\ignore{
In Section \ref{sec:discussion}, we refer to a CDN market trend that makes \DNSUD{} more attractive.
}
\ignore{%%%%%%%
In Section \ref{sec:beyond}, we discuss ways to expand the method to include communication between clients and resolvers, i.e., updating stale records in browser caches and other client applications before the corresponding TTL has expired.
}%%%%%%%
\ignore{
In Section \ref{sec:extension} an extension is presented to provide fast recovery when an authoritative server is unavailable.
}
Finally, conclusions are given in Section \ref{sec:conclusion}.
\ignore{Our analysis indicates that more than $80\%$ of the $50$ most popular domains and the $450-500$ most popular, change their IP address less than once a month, and more than $50\%$ less than once a year. Domain names with short TTL values (less than $10$ minutes) benefit from \DNSUD, by using much larger TTL values and reducing their DNS traffic and the average latencies experienced by clients. Also, \DNSUD{} could ensure a constant and quick recovery from IP address changing for domain names with large TTL values.}  For a detailed version please see \cite{arielThesis}.
%%%%%%%%%%%%%%%%%%%%%%%%%%%%%

%%% Local Variables:
%%% mode: latex
%%% TeX-master: "cwc"
%%% End:

%% file: background.tex
\section{Related work}
\label{sec:background}

\ignore{%%%
\paragraph{Background}
}%%%

\ignore{%%%
\subsection{TTL values}
%\label{sec:}
%%%%%%%%%%%%%%%%%%%%%%%%%%%%%
TTL values of DNS records determine cache durations. Therefore, they affect latency, availability and reliability of the DNS network. Some previous studies modeled caches
as a function of their TTL and some researches provided recommendations for TTL values. 

Previous studies have presented alarming findings about the TTLs selected on websites. They showed the negative effects in choosing low TTLs. \newline
For example: "Cache Me If You Can" showed that longer TTLs have important performance benefits, since caching greatly reduces latency, even more than anycast, as well as reducing traffic. Their scans of deployed DNS show that operators today have little consensus on typical TTLs. 
}%%%

%%%%%%%%%%%%%%%%%%%%%%%%%%%%%
\ignore{%%%
\subsection{DNS packets}
%\label{sec:}
%%%%%%%%%%%%%%%%%%%%%%%%%%%%%
RFC 1035 describes in detail what DNS messages look like. For the purpose of this study, all DNS packets have the following structure:
+---------------------------------+ \newline
| Header\hspace{1.7cm}| \newline
+---------------------------------+ \newline
| Question\hspace{1.45cm}| Question for the name server \newline
+---------------------------------+ \newline
| Answer\hspace{1.55cm} | Answers to the question \newline
+---------------------------------+ \newline
| Authority\hspace{1.28cm} |  Points toward authoritative name server \newline
+---------------------------------+ \newline
| Additional\hspace{1.19cm} | Additional information about the answer \newline
+---------------------------------+ \newline

The header describes the type of packet and determines which fields are contained in the packet. 
}%%%
%%%%%%%%%%%%%%%%%%%%%%%%%%%%%

\ignore{%%%
\subsection{DNS record types}
%\label{sec:}
%%%%%%%%%%%%%%%%%%%%%%%%%%%%%
Although the main function of DNS is to translate domain names into IP address, there are many types for a DNS record. \newline
Figure 2 shows us the most common types.

\begin{figure}[h]
    \centering
    \includegraphics[width=8cm]{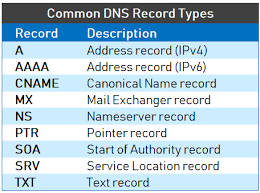}
    \caption{DNS record types}
    \label{fig:dns-types}
\end{figure}
}%%%

%%%%%%%%%%%%%%%%%%%%%%%%%%%%%
\ignore{%%%
\subsection{Incidents in DNS records}
%%%\label{sec:}
%%%%%%%%%%%%%%%%%%%%%%%%%%%%%
Several cases where the wrong IP address has been distributed for a record or where the authoritative servers have become inaccessible, damage was caused to the Internet service's availability:

\begin{itemize}[leftmargin=0.15in]
    \item Microsoft\cite{azure} - In 2019 Microsoft services like Azure, Microsoft 365 umbrella, Dynamics, and DevOps were down at least 2 hours due to an incorrect DNS update. Microsoft explained that: "During the migration of a legacy DNS system to Azure DNS, some domains for Microsoft services were incorrectly updated."
    \item SiteGround\cite{siteground} - In 2021 SiteGround services were not accessible to Google crawlers for several hours due to an incorrect DNS update. The company explained: "We are glad to inform you that we have implemented a fix for the Google bot crawling issue experienced by some sites... Please allow a few hours for the DNS changes to take effect".
\end{itemize}

There were many similar incidents of wrong domain names updates such as Zoho\cite{zoho}, Slack\cite{slack}, Windows Update\cite{win-update}. Common to all cases is the long time it took to solve the problem and return the website to its normal availability. In these cases, clients were instructed to remain patient and wait a few hours for a DNS update. }%%%
%%%%%%%%%%%%%%%%%%%%%%%%%%%%%
\ignore{\subsection{Related work}}
%%%\label{sec:}
%%%%%%%%%%%%%%%%%%%%%%%%%%%%%
Various existing resolvers provide cache flushing services through a web interface, such as Google DNS's "Flush Cache" \cite{flush-google} and Cloudflare's "Purge Cache" \cite{purge-cloudflare}. The interface allows to purge one domain at a time with several seconds of delay between successive purges to limit DDoS attacks. 
Two problems arise; first, Internet service owners must manually flush records in each resolver separately, hoping that each resolver implements cache flushing services; secondly, anyone can delete DNS records from these resolver's cache. \DNSUD{} is somewhat similar, but it is a reliable method that automates the service, providing it to any authoritative and/or resolver.

In \cite{mitigating-dos} Ballani et.al. suggest using the IP address of a domain even if its TTL has expired when the corresponding authoritative server is unreachable due to e.g., a DDoS attack. 
\DNSUD{} provides a similar remedy by enabling the use of much larger TTL values. 
However, unlike \DNSUD{}, \cite{mitigating-dos} does not help if an authoritative server crashes while an erroneous record has been distributed and might make things worse, i.e., increasing the potential damage of DNS cache poisoning attacks \cite{cache-injection-study, poison1, poison2} % , poison3}. 

\nsdionly{
In \cite{moura2018cache} Moura et.al. describe methods used by DNS resolvers to mitigate DDoS attacks, such as using multilevel caching, overriding TTL values, and resending DNS requests when on a timeout. These reduce the impact of DDoS attacks, but unlike \DNSUD{}, \cite{moura2018cache} does not help to quickly recover from an erroneous update of a domain name.
}

The "DNS push Notifications" RFC \cite{rfc8765}, suggests a mechanism wherein an authoritative server asynchronously notifies all its clients of changes to its DNS records via a TCP connection.
This mechanism has a fundamental drawback:
Each authoritative server is required to maintain a persistent TCP connection with all of its clients to facilitate updates.
Consequently, it might be suitable for an IoT product communicating with its designated vendor, but not for client software, such as browsers, which cannot maintain TCP connections with a large number of websites simultaneously only to await updates. 
Furthermore, this mechanism generates a significant load on the authoritative server and necessitates more substantial modifications in the DNS than those required by \DNSUD.

%%%%%%%%%%%%%%%%%%%%%%%%%%%%%
%%% Local Variables:
%%% mode: latex
%%% TeX-master: "cwc"
%%% End:

%% file: solution.tex
%\newpage \newpage
\section{\DNSUD{}}
\label{sec:solution}
%%%% \subsection{Overview}
\label{sec:overview}
%%%%%%%%%%%%%%%%%%%%%%%%%%%%%
\ignore{%%%%%%%%%%
At a high level, the method has three steps: first, resolvers get a list of domains whose records have been recently updated; secondly, resolvers remove any record that appears in the list from their cache; and thirdly, upon receiving a request for a removed domain, a cache miss occurs, and the resolver performs a new resolution in the standard DNS system. In this resolution the resolver gets the updated IP address from the corresponding authoritative server. 
}%%%%%%%%%

At a high level, the method has three steps:
\begin{description}%%[leftmargin=0.12in]\setlength\itemsep{0.25em}
%%    \item [step 1] Resolvers are provided with a list of domains whose records have been recently updated.
    \item [Step 1] The \DNSUD{} service repeatedly supplies resolvers with a worldwide list of domains whose record (IP address) has been recently updated.
    \item [Step 2] Each resolver removes from its cache any record that appears in the last list it received.
    \item [Step 3] Upon receiving a query for a removed domain, a cache miss occurs, and the resolver performs a new resolution in the standard DNS system getting the updated IP address from the corresponding authoritative server. 
\end{description}

\begin{figure}[h!]
    \centering
    \includegraphics[width=7cm]{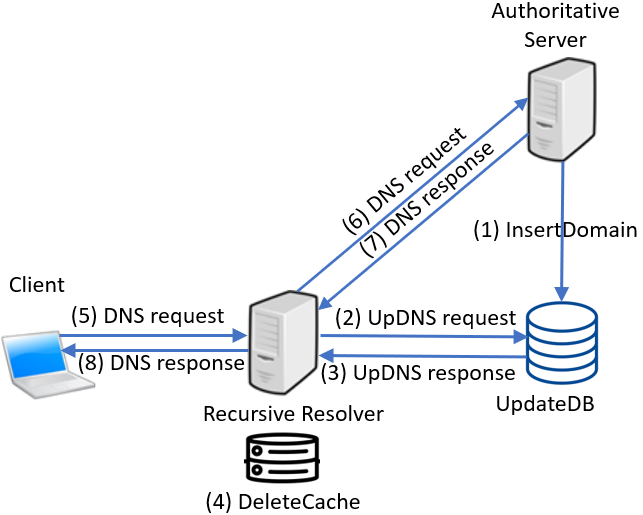}
    \caption{\DNSUD{} system overview}
    \label{fig:dnsud-pure}
\end{figure}

\begin{figure}[h!]
    \centering
    \includegraphics[width=7cm]{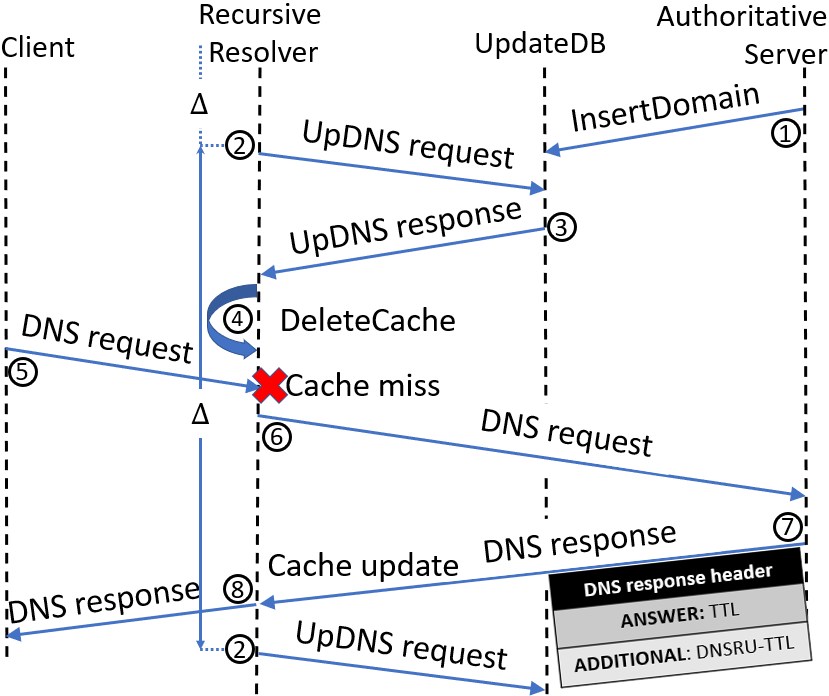}
    \caption{\DNSUD{} events and messages flow: \newline
    1. InsertDomain - Upon updating a record, the authoritative server sends the new record to the \UpdateDB. \newline 
    2. UpDNS request - The recursive resolver queries the \UpdateDB{} every $\Delta$ time units.  \newline
    3. UpDNS response - The resolver receives the list of domains that have been updated in the last $\Delta$ time.  \newline
    4. DeleteDNS - The recursive resolver deletes the domains in the list from its local cache. \newline
    5. A cache miss occurs when a client sends a request to resolve a domain name that has been recently updated. \newline
    6. The recursive resolver issues a request to resolve the domain in the standard DNS system.  \newline
    7. The DNS system resolves the request with two different TTLs in a backward compatible way (Section \ref{sec:new-ttl}). \newline
    8. The recursive resolver inserts the new domain name into its cache and sends it to the client.}
    \label{fig:dnsud}
\end{figure}

\DNSUD{} maintains a global database, called \UpdateDB{}, that contains a list of domain names that have been recently updated, see Figure \ref{fig:dnsud-pure}. Each high-level step is implemented as follows;
First, upon updating a domain name record, an authoritative server sends a message to the \UpdateDB{} to insert the updated record into the \UpdateDB.
Secondly, resolvers query the \UpdateDB{} every $\Delta$ time units (e.g., $\Delta=1_{min}$) for the list of domain names that have been updated in the last $\Delta$ time units.
Finally, each resolver deletes from its cache each domain name that appears in the list it has received. In a subsequent request to such a domain, due to a cache miss, the resolver resolves the domain name in the standard DNS system, obtaining a fresh record from the corresponding authoritative server.

\ignore{%%%%%%%%%%%%%%%%%%%%%%%%%%%%%%
%tbhp h
\begin{figure*}[!b]
    \centering
    \includegraphics[width=\textwidth]{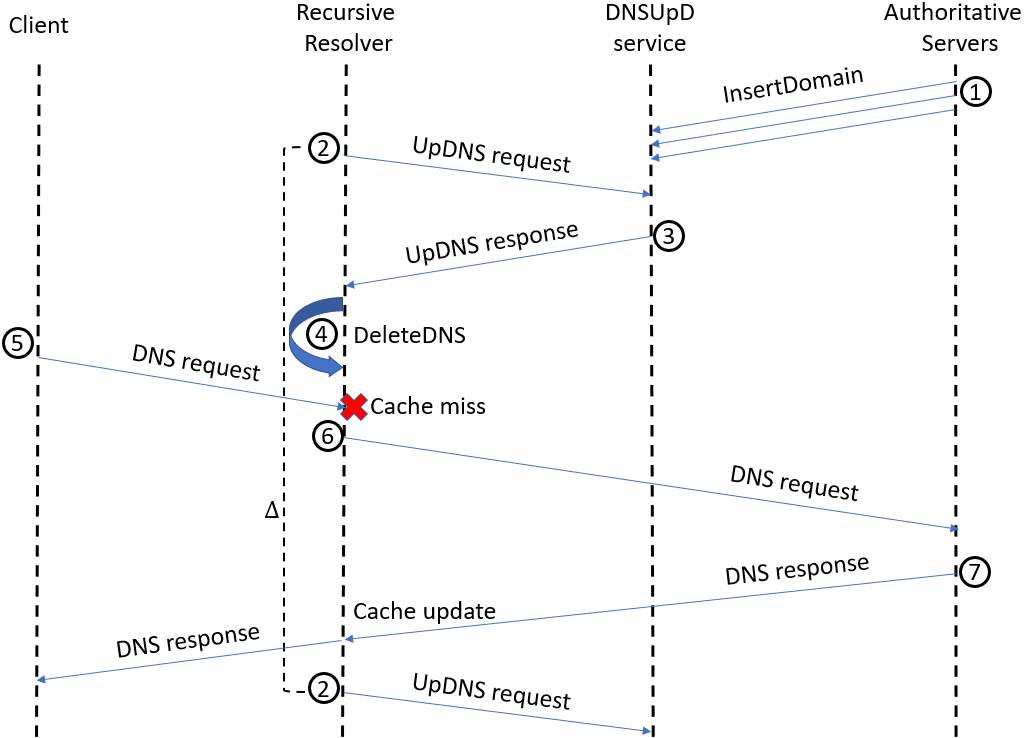}
    \caption{\DNSUD{} system flow chart: \newline
    1. InsertDomain - Upon updating a record, authoritative servers send the new record to the \UpdateDB. \newline 
    2. UpDNS request - The recursive resolver queries the \UpdateDB{} every $\Delta$ time units.  \newline
    3. UpDNS response - The recursive resolver receives the list of domains that have been updated in the last epoch.  \newline
    4. DeleteDNS - The recursive resolver deletes the domains in the list from its local cache. \newline
    5. DNS request - A cache miss occurs when a client sends a request to resolve a domain name that has been recently updated.  \newline
    6. DNS request - The recursive resolver resolves the request in the standard DNS system. \newline
    7. DNS response - The DNS system resolves the request. The recursive resolver inserts the new domain name into its cache and sends it to the client.}
    \label{fig:dnsud}
\end{figure*}
}%%%%%%%%%%%%%%%%%%%%%%%%%%%%%%%

\ignore{%%%%%%%%%%%%%%%%%%%%%%%%%%%%%%%%%%%%%%%%%%%%%%%%%%%%%%%%%%%%%%%
The idea of \DNSUD{} method is very simple; we add a server with a database that contains a list of DNS records that are outdated. When updating a DNS record, the owner of authoritative server updates the record in their own server, as well as our server. 
The resolvers access our server periodically and receive the newest list of domains, updated. Then, they delete each domain in their local cache (if it exists in the cache). Now, next time the resolver receives a request, it forwards the request to the DNS network for the updated record.}

\ignore{In Section, \ref{sec:IPupdate} we discuss an extension of the method. In this extension, the recursive resolver resolves a domain name directly from the \UpdateDB{}. Thus, \UpdateDB{} stores not only the recently updated domain names but also the updated IP addresses for them. This extension enables IP updates for DNS records even if their corresponding authoritative servers are inaccessible.}
%%%%%%%%%%%%%%%%%%%%%%%%%%%%%

%\subsection{\DNSUD{} two TTLs}
\subsection{\DNSUD{}: Adding a large TTL}
\label{sec:increase-ttl}
%%%%%%%%%%%%%%%%%%%%%%%%%%%%
\ignore{%%%%%%%%%%%%%%%%%%%%%
\DNSUD{} enables Internet services to significantly increase the TTL values associated with their domain names, on top of the benefits described in Section \ref{sec:solution}.
TTL increasing is possible due to the short period from the time an authoritative sends an InsertDomain request to the \UpdateDB{} until any client gets the updated record information in response to its DNS request (around one minute), see Section \ref{sec:recovery-time}.}%%%%%%%%%%

Since \DNSUD{} ensures a fast recovery time for domains using it, regardless of the TTL value, the use of larger TTL values is no longer limited.
Therefore, we associate a second TTL value, called \DNSUD-TTL, with the records of domains that use \DNSUD{}.
\ignore{%%%%%%%%%%%%%%%%%%%%%%%%%%%%%%%%%%%%%%%%%%%
\DNSUD{} enables Internet services to significantly increase the TTL values associated with their domain names, on top of the benefits described in Section \ref{sec:solution} due to the short recovery time (around one minute) achieved by \DNSUD, see Section \ref{sec:recovery-time}.}%%%%%%%%%%%%%%%%%%%%%%%%
%%%%%%%%%%%%%%%%%%%%%%%%%%%%
%%\subsection{Backward compatibility of the \DNSUD{}}
\label{sec:new-ttl}
Notice that it is still necessary to maintain the shorter TTL value for communication between resolvers and clients, as well as for backward compatibility and gradual adoption of \DNSUD{}.
However, the standard DNS does not support the association of an additional TTL value with DNS replies and records.
\ignore{%%%%%%%%%%%%%%%%%%%%
Thus, we need to find a way for an authoritative server to send a different TTL value for the resolvers that use the \DNSUD{} method without influencing resolvers and clients that still do not use the service.
To enable backward compatibility with the current DNS system, we should describe how some of the resolvers can use the \DNSUD{} service while the others operate as usual.} %%%%%%%%%%%%%%%%%%
To address this mismatch, we utilize the optional ADDITIONAL section in DNS replies and records \cite{rfc1035} to include the \DNSUD-TTL value.
The maximum possible value for \DNSUD-TTL is 2,147,483,647 seconds \cite{rfc1034} (equating to more than 68 years), can be assigned as the \DNSUD-TTL, see Section \ref{sec:recovery-time}.
\ignore{%%%%%%%%%
To add the \DNSUD-TTL value while maintaining the traditional TTL value in the authoritative responses, in a backward compatible way, the optional ADDITIONAL section \cite{rfc1035} is used for the \DNSUD-TTL value.
Within the ADDITIONAL section, a new record with the new TTL value - the "\DNSUD-TTL" is added.
}%%%%%%%%%
For example the DNS answer sent by a Google authoritative server would be: \\
\textbf{ANSWER section:} \newline
google.com {} {} {} {} $600$ {} {} {} {} IN {} {} {} {} A {} {} {} {} $172.253.115.100$ \newline
\textbf{ADDITIONAL section:} \newline
\DNSUD.google.com $2,147,483,647$ {} IN A {} $172.253.115.100$ \\

To implement the \DNSUD-TTL functionality, Internet service owners need to include the "\DNSUD-TTL" in their DNS zone file \cite{zone-file}. 
Clients and resolvers that do not implement \DNSUD{} disregard this ADDITIONAL section and retrieve the traditional TTL value from the ANSWER section. 
While resolvers that support \DNSUD{} extract the "\DNSUD-TTL" from the ADDITIONAL section.
This approach enables an authoritative server to communicate with both resolvers that have implemented the \DNSUD{} method and those that have not.

\ignore{%%%%%%
Internet service owners add the "\DNSUD-TTL" in their DNS zone file \cite{zone-file}. 
Clients and resolvers that do not use \DNSUD{} ignore the ADDITIONAL section and take the traditional TTL from the ANSWER section, while resolvers that use \DNSUD{} take the "\DNSUD-TTL" for the domain from the ADDITIONAL section.
Thus, an authoritative server that uses \DNSUD{} can communicate with resolvers that do, or do not, implement the method.
}%%%%%%
%%%%%%%%%%%%%%%%%%%%%%%%%%%%%%%%%%%%%%%%%%%%%%%%%%%%%%%%%%%%%%%%%%%%%%%%%%%%

%%% Local Variables:
%%% mode: latex
%%% TeX-master: "cwc"
%%% End:

%% file: measurements.tex
%\newpage
\section{Measurements and Evaluation}
\label{sec:measurements}
\nsdionly{%%%%%%%%%%%%%%%%%%%%%%%%%%%%%%%
In this section, we address three critical challenges:
First, we verify the applicability of the method for the majority of domain names.
Second, we determine the value for $\Delta$ by observing the current TTL values.
Finally, we assess the existing volume of futile traffic to illustrate the potential traffic savings for domain owners.
Addressing these challenges will provide valuable insights and data to support the adoption of \DNSUD{} and its integration into the DNS ecosystem.
}%%%%%%%%%%%%%%%%%%%

In this part, we address three critical challenges:
First, we verify the applicability of the method for the majority of domain names.
Secondly, we determine the value for $\Delta$ by observing the current TTL values.
Finally, we assess the existing volume of futile traffic to illustrate the potential traffic savings for domain owners.
\ignore{
\DNSUD{} implementation details are given in Appendix \ref{sec:drill-down}.
}

%%%%%%%%%%%%%%%%%%%%%%%%%%%%%%%%%%%%%%%%%%%%%%%%%%%%%%%
\subsection{Percentage of domains suitable for \DNSUD{}}
\label{sec:freq-updates}

We classify domain names into two categories based on their frequency of IP address changes: \emph{stable} and \dynamic.
The average time between IP address changes for a domain $D$ is denoted as $T_{update}^D$.
For stable domains,  $T_{update} > 1$ month, while other domains are deemed \dynamic.
All stable domains are considered suitable for \DNSUD.
\ignore{%%%%%%%%%%%%%%%%%%%%%%%%%%%%%%%%%%%%%%%%%
We ensure the method's applicability and benefits for both types, with $T_{update}^D$ denoting the average time between IP address changes for a domain $D$.

Here we verify \DNSUD{} relevance for the majority of domain names.
We need to ensure that the method is applicable and beneficial for a significant number of domains across the Internet.
We categorize Internet domain names into two types: 
\emph{stable} domains which change their IP address less than once per month on average, and \dynamic{} domains, which frequently change their IP address (e.g., Akamai clients).
Let $T_{update}^D$ be the average time between IP address changes for domain $D$. That is, $T_{update} > 1$ month for the \emph{stable} domains.
}%%%%%%%%%%%%%%%%%%%%%%%%%%%%%%%%%%%%%%%%%%%%%%%%
\ignore{%%%%%%%%%
\dynamic{} domains include some CDNs \cite{kazunori2012cdn, centralized-dns}, such as Akamai, but not all CDNs fall into this category, as discussed below.
We measured the frequency of IP address changes for many domains and found that more than $80\%$ are \emph{stable} domains.
}%%%%%%%%%

To identify and estimate the number of \emph{stable}  domains, we need to collect the historical data of $T_{update}$.
We used SecurityTrails \cite{securitytrails}, a paid service, to collect data on domain IP address changes over the last $3$ to $15$ years.
To confirm the sensitivity of their sampling frequency - ensuring domains are not updating their IP addresses back and forth without SecurityTrails detection - we sampled the IP addresses of the examined domains every five minutes for a month.
For domains with $T_{update} > 1$ month according to SecurityTrails, our measurements did not identify any outliers domains.
However, for domains with $T_{update} < 1$ month according to SecurityTrails, we noticed a few IP changes that were missed.

\begin{figure}[ht]
    \centering
    \includegraphics[width=8.5cm]{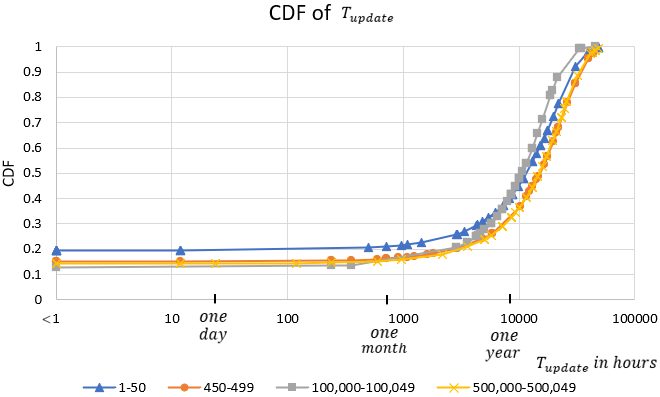}
    \caption{CDF of $T_{update}$ in hours.
    The x-axis of each point represents the $T_{update}^D$ for domain name $D$ while the y-axis represents the CDF. \\
    }
    \label{fig:cdf}

    \includegraphics[width=5cm]{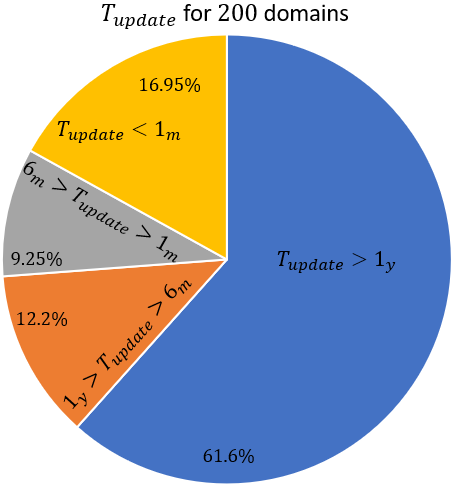}
    \caption{$T_{update}$ for $200$ domains. $m = month$, $y = year$. \\
    $T_{update}^D > 1$ year for $61.6\%$.
    $T_{update}^D > 1$ month for $83.05\%$.
    %%%%%%%%%%%%%%%%%%%%%%%%%%%%%%%%%%
    \ignore{
    For domain D:
    The probability that $T_{update}^D > 1$ year is $61.6\%$. \\
    The probability that $T_{update}^D > 1$ month is $83.05\%$
    }%%%%%%%%%%%%%%%%%%%%%%%%%%%%%%%%%%%%%
    }
    \label{fig:domains-update}
\end{figure}

We obtained a list of domains sorted by popularity (including sub-domains) from MOZ and Alexa \cite{top-domains, top-domains2} and sampled $200$\footnote{SecurityTrails licensing enabled to get only $200$ $T_{update}$ measurements.}, as follows: the up-to-date top $50$ and the top $450-499$ of MOZ, and top $100,000-100,049$ and top $500,000-500,049$ of Alexa from $2021$.
Then, we measured $T_{update}^D$ for each of these domains. We did not measure the entire $600,000$ domains due to SecurityTrails licensing constraints. As can be seen in Figure \ref{fig:cdf}, the distribution of $T_{update}$ is quite consistent across the four ranges.
For domains that were updated at least once a day, we took their $T_{update}$ from our samplings rather than from SecurityTrails.
Other papers that use the historical data on domain IP address changes collection of SecurityTrails are \cite{Sectrails1, Sectrails2, Sectrails3}.

%%%%%%%%%%%%%%%%%%%%%%%%%%%%%%%%%%%%%%%%%%%%%%%%%%%%%%%%%%%%%%%%%%%%%%%%%%%%%%%%%%%%%%%
\ignore{%%%%%%%%%%%%%%%%%%%%%%%%%%%%%%%%%
\begin{itemize}[leftmargin=0.15in]\setlength\itemsep{0.25em}
    \item $T_{update}$ > $1$ month - for $80\%$ of domains among the top $50$ and $85\%$ among the last $50$. 
    \item $T_{update}$ > $6$ months - for $70\%$ of domains among the top $50$ and $75\%$ among the last $50$. 
    \item $T_{update}$ > $1$ year - for $60\%$ of domains among the top $50$ and $65\%$ among the last $50$.
\end{itemize}
}%%%%%%%%%%%%%%%%%%%%%%%%%%%%%%%%%%%%
\textbf{Observations:} \\
The $T_{update}$ measurements are summarized in Figures \ref{fig:cdf} and \ref{fig:domains-update}, indicating that the group of \emph{stable} domains constitutes  more than $80\%$ of the examined domains. 

We also investigated whether domains serviced by CDNs are inherently \dynamic{} domains. 
\ignore{%%%%%%%%%%%%%%%%%%%%%%%%%%%%%%%%%%%%%%%%%%%%%%%%%%%%%%%%%%%%%%%%%%
We used the statistics of CDNs among the websites: $24\%$ of the websites use common reverse proxy services \cite{proxy}. Among them, Cloudflare is used by $79.8\%$, Fastly by $6.9\%$, and Akami by $1.8\%$.
}%%%%%%%%%%%%%%%%%%%%%%%%%%%%%%%%%%%%%%%%%%%%%%%%%%%%%%%%%%%%%%%%%%%%%%
The update results for these CDNs (Akamai, Cloudflare, and Fastly) are as follows:
\begin{itemize}%%[leftmargin=0.12in]\setlength\itemsep{0.25em}
    \item Akamai's domains are updated frequently. Naturally, Akamai's domains being \dynamic{} domains, have no choice but to provide short TTL values, around tens of seconds. According to market share statistics, Akamai is used by about $1\%$ of the total Internet services \cite{proxy}.
    \item $90\%$ of Cloudflare's domains are \emph{stable}.
    The average $T_{update}$ of these \emph{stable} domains is $20$ months, while Cloudflare sets their TTL to $300$ seconds.
    \item All of Fastly's domains are \emph{stable}.
    The average  $T_{update}$ for these \emph{stable} domains is almost $2$ years, while the TTL of $80\%$ of them is less than $10$ minutes. Meanwhile, the TTL of the remaining $20\%$ is $1$ hour.
\end{itemize}
Surprisingly most of Cloudflare's and Fastly's CDN serviced domains are \emph{stable}. This suggests that some domains serviced by CDNs are potential users of \DNSUD.

\textbf{Subdomains:} \\
\ignore{
Note that while a top-level domain is \emph{stable}, (\dynamic) some of its subdomains may be \dynamic{} (\emph{stable}).
We further observed that among the $200$ domains we measured from the SecurityTrails database, some were subdomains.
Below are a few examples of domains and subdomains of differing types:
}
Note that while a top-level domain might be \emph{stable}, some of its subdomains could be \dynamic, and vice versa.
A few subdomains are among the $200$ domains we measured.
Two examples of domains and subdomains of opposing types are:
\begin{itemize}%%[leftmargin=0.12in]\setlength\itemsep{0.25em}
    \item Domain office.com is \emph{stable}, with $T_{update}^{office.com} = 3$ years. It has \emph{stable} subdomains, such as parteners.office.com, excel.office.com, and dev.office.com, but also \dynamic{} subdomains like support.office.com and blogs.office.com.
    \ignore{%%%%%%%%%%%%%%%%%%%%%%%%%%%
    \item ieee.org is a \emph{stable} domain, with $T_{update}^{ieee.org} = 33$ months. It has some \emph{stable} subdomains, such as standards.ieee.org, ewh.ieee.org, and attend.ieee.org, but also an \dynamic{} subdomain serviced by Akamai - ieeexplore.ieee.org.
    \item adobe.com is \dynamic{} domain, with a \emph{stable} subdomain serviced by Fastly - stock.adobe.com.
    }%%%%%%%%%%%%%%%%%%%%%%%%%%%%%%%%%
    \item Domain netflix.com is \dynamic{} domain, but its media.netflix.com and dvd.netflix.com are \emph{stable} subdomains.
\end{itemize}

\ignore{%%%%%%%%%%%%%%%%%%%%%%%%%%%%%%%%%%%%%%%%%%%%%
Essentially some websites that are frequently updating their records need minor modifications to become "stable domains". For example, Facebook's DNS response has a primary IP address, which sometimes updates to its secondary IP address, probably for maintenance. Then, it switches back to its primary IP address after a short time. Facebook may consider changing its behavior to using two constant IP addresses in the same DNS response like Amazon. With this method, if the first IP address is not accessible, clients try to access the second one.
}%%%%%%%%%%%%%%%%%%%%%%%%%%%%%%%%%%%%%%%%%%%%%%%%%%%
%%%%%%%%%%%%%%%%%%%%%%%%%%%%%%%%%%%%%%%%%%%%%%%%%%%%%%%%%%%%
\subsection{Determining $\Delta$}
\label{sec:ttl-measurments}
\begin{figure}[h]
    \centering
    \includegraphics[width=5cm]{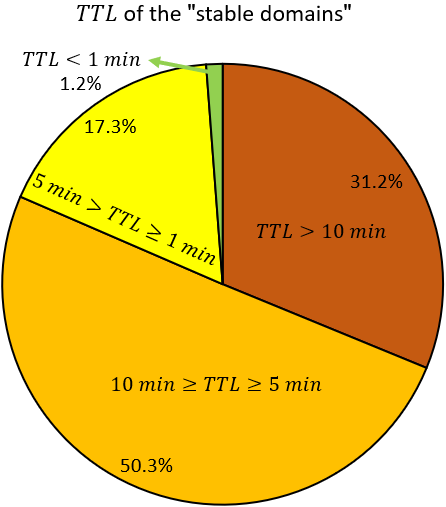}
    \caption{Distribution of the TTLs of the \emph{stable} domains\ignore{:\\
    $31.2\%$ of the domains have TTL $>$ 10 minutes. \\
    $81.5\%$ of the domains have TTL $\geq$ 5 minutes. \\
    $98.8\%$ of the domains have TTL $\geq$ 1 minute.}}
    \label{fig:ttl-infreq}
\end{figure}

$\Delta$ in \DNSUD{} is roughly the time it would take for an IP update to disseminate to the relevant resolvers, which we refer to as the recovery time in Section \ref{sec:increase-ttl} and formally define in Section \ref{sec:recovery-time}.  
In the present DNS system the recovery time for a domain is its TTL value.  
To ensure that the recovery time of domains in \DNSUD{} does not exceed their recovery time in the current DNS system, we set $\Delta$ to the first percentile of the TTL values of \emph{stable} domains.
As observed in Figure \ref{fig:ttl-infreq}, the  $1.2$th percentile of the TTL values of \emph{stable} domains is $1$ minute.
Hence, we set $\Delta=1_{min}$. 

\ignore{%%%%%%%%%%%%%%%%%%%%%%%%%%%%%%%%
the smallest TTL values associated with \emph{stable} domains 
While our method introduced a new much larger TTL value, \DNSUD-TTL, $\Delta$
is determined based on the minimal desired recovery time for domain owners. 
This is because the recovery time in the \DNSUD{} system primarily depends on $\Delta$, as discussed in Section \ref{sec:recovery-time}.
 
Currently, domain owners choose their TTL value as the period of time after which their domains should be purged from caches worldwide.  
Thus we determine $\Delta$ by examining the smallest TTL values associated with \emph{stable} domains.
Hence in Figure \ref{fig:ttl-infreq} we extracted the TTL values of the aforementioned $200$ websites.
Only $1.2\%$ of the \emph{stable} domains have a TTL of less than $1$ minute, while the remaining have a larger TTL value. Therefore, we determine $\Delta=1_{min}$. 
}%%%%%%%%%%%%%%%%%%%%%%%%%%%%%%%%%%%%%%
 
\ignore{%%%%%%%%%%%
Note that the TTL for $31.2\%$ of the \emph{stable} domains exceeds $10$ minutes, indicating that a significant number of websites avoid using short TTLs.}%%%%%%%%%%%%
\nsdionly{Notice, that the TTL of $68.8\%$ of the \emph{stable} domains is less than $10$ minutes and thus their recovery from an erroneous configuration may take up to $10$ minute. The current recovery time of the other $31.2\%$ equals their TTL which is larger than $10$ minutes.}
\ignore{%%%%%%%%%%%%%%%%%%%%%%%%%%%%%%%%%%%%%%%%%%%%%%%%%%%%%
While the former suffer the drawbacks of short TTLs (higher load on their authoritative), the latter enjoys the benefits of long TTLs, see Section \ref{sec:intro}.}%%%%%%%%%%%%%%%%%%%%%%%%%
%%%%%%%%%%%%%%%%%%%%%%%%%%%%%%%%%%%%%%%%%%%%%%%%%%%%%%%%%%%%
\subsection{Futile traffic}
\label{sec:ttl-unchanged-domains}
Here we assess the current amount of futile traffic in order to demonstrate potential traffic savings to authoritative servers, incentivizing them to adopt \DNSUD.
This assessment quantifies the volume of unnecessary or redundant traffic generated because of the short TTL values.
We assume an ideal cache model in which DNS records are not evicted from the cache until the corresponding TTL has expired, i.e., ignoring eviction due to full cache, resolver crashes, and resolvers manipulations \cite{moura2018cache, resolver-manipulations}.
We noticed that fewer than $0.2\%$ of DNS responses contain a DNS record that is different than the previous record received.
Increasing the TTL values, as facilitated by \DNSUD{}, would significantly reduce futile traffic, and under the ideal cache assumption, this traffic could be completely eliminated.
\ignore{%%%%%%%%%%%%%%%%%%%%%%%%%%%%%%%%%%%%%%%%%%%%
In Section \ref{sec:freq-updates}, we consider two types of domain names: "repetitive domains", most of them are changing their IP address every day, and "stable domains" that change their IP address less than once per month on average.}%%%%%%%%%%%%%%%%%%%%%%%%%%%%%%%%%%%%%%

More specifically, let $traffic\text{-}ratio^D = \frac{T_{update}^D}{TTL^D}$ for domain $D$ be the ratio of the average time between IP address changes for domain $D$ and its corresponding TTL value. 
Theoretically the lower bound on $traffic\text{-}ratio^D$ occurs when the TTL expires exactly when an IP address is changed, i.e., $\Omega(traffic\text{-}ratio^D) = 1$.
In this case, there would be no futile traffic. 
We consider $traffic\text{-}ratio^D$ to be its futile traffic.

\begin{figure}[h]
    \centering
    \includegraphics[width=9cm]{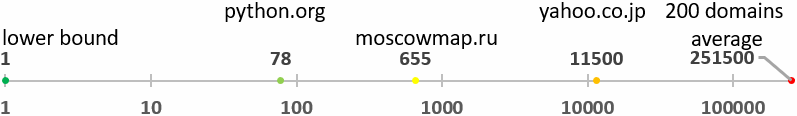}
    \caption{$traffic\text{-}ratio$ of the \emph{stable} domains on a logarithmic scale.
    The $traffic\text{-}ratio$ is considerably larger than the lower bound, which equals $1$.}
    \label{fig:ex-traffic}
\end{figure}

In Figure \ref{fig:ex-traffic}, we computed the futile traffic, which is the ratio mentioned above, using the $T_{update}^D$ and TTL values from Sections \ref{sec:freq-updates} and \ref{sec:ttl-measurments}.

Notice that 'python.org' and 'moscowmap.ru' have the lowest ratios, while 'yahoo.co.jp' has the lowest ratio among those with a short TTL value (less than $10$ minutes). 
The average $traffic\text{-}ratio$ for the $200$ domains is quite large, at $251,500$. 
The highest observed ratio is $2,365,200$, found at 'amazon.in'.

\nsdionly{
\begin{itemize}[leftmargin=0.12in]\setlength\itemsep{0.25em}
    \item The smallest ratio is $78$ for "python.org", which has a unique TTL value of $7$ days even though its IP address is updated every $16$ months on average. Even this futile traffic is far from the lower bound. If we exclude it (due to its unusually high TTL), the next smallest ratio is $655$ for "moscowmap.ru".
    \item The minimum ratio for domain names with a short TTL values (less than $10$ minutes) is $11,500$, observed in "yahoo.co.jp". It maintains a TTL value of $300$ seconds, yet its IP address is updated, on average, once a month.
    \item The average $traffic\text{-}ratio$ is $251,500$. The highest is $2,365,200$, observed in "amazon.in". This domain has a TTL value of $60$ seconds despite its IP address being updated once every $4.5$ years, on average.
\end{itemize}
}

Notice that the reported $traffic\text{-}ratio$ with the ideal cache assumption is a lower bound; Evicting a domain from the cache earlier, increases the $traffic\text{-}ratio$.
As noted above, with \DNSUD{} the futile traffic in an ideal cache model can be even eliminated by choosing an immense TTL.

%%%%%%%%%%%%%%%%%%%%%%%%%%%%%

%%% Local Variables:
%%% mode: latex
%%% TeX-master: "cwc"
%%% End:

%% file: analysis.tex
%\newpage
\section{Analysis}
\label{sec:analysis}

Here the above measurements are used to analyze (A) The recovery time of a domain, (B) the load on the authoritative servers, (C) feasibility and scalability of \DNSUD{} and finally (D) assessing the economic implications.
\ignore{
\DNSUD{} implementation details are given in Appendix \ref{sec:drill-down}.
}

\ignore{%%%%%%%%%%%%%%%%
In this section we use the above measurements to analyze key parameters of the \DNSUD{} method:
(A) The recovery time of a domain, (B) the load on the authoritative servers, (C) feasibility and scalability of \DNSUD{} and finally (D) assessing the economic implications.\\
}%%%%%%%%%%%%%%%%

\ignore{%%%%%%%%%%%%%%%%%%
One may wonder, how fast is the recovery in \DNSUD{} from an erroneous domain update? and, what is the load on the authoritative servers in \DNSUD{}? Is the suggested method feasible? And finally, what are the economic implications?
}%%%%%%%%%%%%%%%%%%%%%%%%%%

\subsection{\DNSUD{}: Domain recovery time}
\label{sec:recovery-time}
Let the recovery time of domain $D$ be the interval of time from when $D$ is updated in its authoritative server, called $AuthS$, until all resolvers have deleted domain record $D$ from their cache.
We consider a domain worst-case recovery time with and without the \DNSUD{}.
With \DNSUD{} it is bounded by $\Delta + l \approx 66$ seconds where $l << \Delta$, independent of the TTL value. 
Without \DNSUD{}, it equals to the TTL value.
Here, we assume an ideal cache model, hence the results are an upper bound on the worst-case recovery time.
Let us define:
\begin{description}%%[leftmargin=0.12in]\setlength\itemsep{0em}
    \ignore{ %%%%%%%%%%%%%%%%%%%%%%%%%%%
    \item [$\mathbf{T_{res}}$:] - The maximum time for a resolver to obtain a fresh record from \UpdateDB. This parameter depends mainly on the frequency($\Delta$) of requests to the \UpdateDB{} and their processing time.
    }%%%%%%%%%%%%%%%%%%%%%
    \item [\textbf{RECOVERY}$^{\textbf{\DNSUD}}$ (\textbf{RECOVERY}$^{\text{DNS}})$:] The worst-case recovery time with (without) \DNSUD. 
    \ignore{The \textbf{recovery time} is defined as the time period from the time a domain name is updated in the corresponding authoritative server until all resolvers have deleted the outdated domain name from their cache.}
    \item [$\mathbf{R^D}$:] The group of resolvers through which clients query $AuthS$ for domain $D$.
    \item [$\boldsymbol{\Delta_{r}}$:] The inter UpDNS-request time for resolver $r\in R^D$ ($=\Delta=1_{min}$). 
    \item [$\mathbf{T^{ID}}$:] The worst-case time to execute the Insert Domain operation, step 1 in Fig. \ref{fig:dnsud}.
    That is, the time since $AuthS$ sends the InsertDomain($D$) request until $T_{step1}$, the time at which \UpdateDB{} inserts the record into its database. $T^{ID}$ is expected to be at most $2$ seconds (typically less than $1$ second) since $AuthS$ sends this request immediately upon updating domain $D$.
    \item [$\mathbf{T_{r}^{UD}}$:] The worst-case time to perform steps 2-4 in Fig. \ref{fig:dnsud}. That is, the time since resolver $r$ sends an UpDNS request, which arrived at the \UpdateDB{} after time $T_{step1}$, until resolver $r$ deletes domain $D$ from its cache. $T_{r}^{UD}$ is expected to be at most $4$ seconds (typically less than $1$ second).
\end{description}
\ignore{%%%%%%%%%%%
We also assume that the TTL value of the domain $D$ is greatly extended after \DNSUD{} usage.}%%%%%%%%%%%%%%

Then:

\begin{center}
 \text{RECOVERY}$^{\text{\DNSUD}} = T^{ID} + \max_{r\in R^D} (\Delta_{r} + T_{r}^{UD}) $
\end{center}
 
This is the sum of three time periods: 
(1) time for the $AuthS$ to perform step 1 in Fig. \ref{fig:dnsud} ($T^{ID}$), 
(2) time for any resolver $r$ to perform steps 2-4 in Fig. \ref{fig:dnsud} ($T_{r}^{UD}$), 
and (3) resolver's $\Delta_{r}$. 
While
\begin{center}
\text{RECOVERY}$^{\text{DNS}} = TTL$ \par
\end{center}

\begin{figure}[h]
    \centering
    \includegraphics[width=6cm]{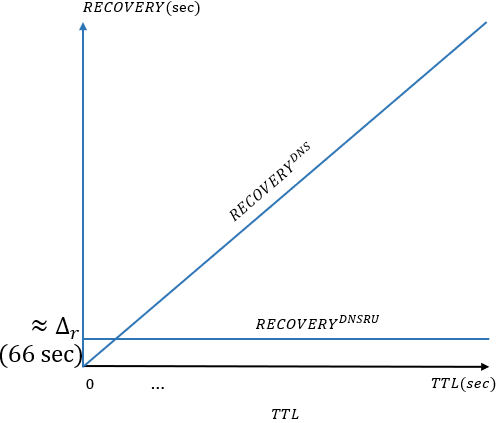}
    \caption{Illustrative diagram: The worst-case recovery time in seconds assumes an ideal cache model before and after \DNSUD{} usage.}
    \label{fig:recovery-dns}
\end{figure}

In Fig. \ref{fig:recovery-dns} we plot \text{RECOVERY}$^{\text{\DNSUD}}$ and \text{RECOVERY}$^{\text{DNS}}$. \\
This is an illustrative plot, \text{RECOVERY}$^{\text{DNS}}$ increases linearly as a function of the TTL, since in the worst-case scenario, some resolver may hold the old record value until the TTL expires. 
While, \text{RECOVERY}$^{\text{\DNSUD}}$ is bounded by $\Delta_{r} + l$ where $\Delta_{r} = 60$ seconds and $l$ is at most $6$ seconds and usually $<2$ seconds.

\ignore{%%%%%%%%%%%%%%%%%%%
\begin{figure}[h]
    \centering
    \includegraphics[width=7cm]{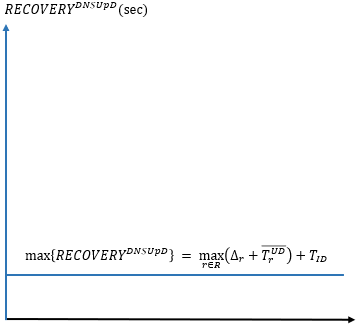}
    \caption{Total recovery time in seconds after \DNSUD{} usage.}
    \label{fig:recovery-dnsud}
\end{figure}
}%%%%%%%%%%%%%%%%%%%%%%%%%%

%%%%%%%%%%%%%%%%%%%%%%%%%%%%%%%%%%%%%%%%%%%%%%%%%%%%%%%%%%%%%%%%%%%%%%%%%%%
\subsection{Load on authoritative servers}
\label{sec:load-dns}
\begin{figure}[h]
    \centering
    \includegraphics[width=7.5cm]{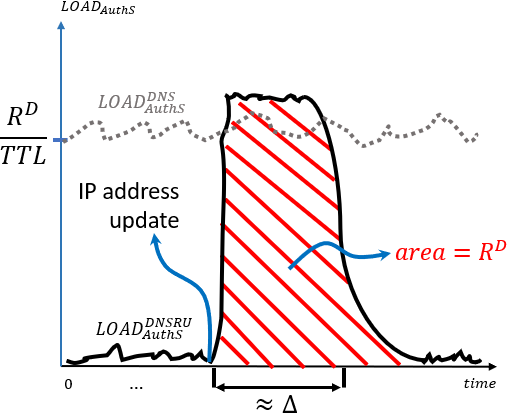}
    \caption{Sketch (illustrative diagram) of $AuthS$ load due to a highly popular domain $D$ under the ideal cache model w/o \DNSUD{}.}
    \label{fig:load}
\end{figure}

\DNSUD{} changes the traffic patterns of authoritative servers that use \DNSUD{}, especially for those with popular domains and have a high $traffic\text{-}ratio$ (Section \ref{sec:ttl-unchanged-domains}).
Let $LOAD_{AuthS}$ be the number of requests per second on an authoritative server $AuthS$ for a popular domain $D$ with a high $traffic\text{-}ratio$ (such as Amazon.com, Baidu.com, Reddit.com, Office.com, eBay.com, etc.).
In the current DNS system, each popular domain is likely to generate a request on $AuthS$ at least once per the domain TTL, for most resolvers $r\in R^D$.
This could result in millions of requests \cite{resolvers} per the domain TTL.
On the other hand, using \DNSUD{}, this load may be significantly reduced because, except when domain $D$ is being updated or evicted, only new resolvers or those recovering from crashes, query $AuthS$ for $D$.
In Figure \ref{fig:load}, we sketch a diagram of the load on authoritative servers for popular domains with and without the \DNSUD{}.
We assume that \DNSUD{}-TTL$^D > T_{update}^D$ and assume an ideal cache model. Let us define:
\begin{description}%%[leftmargin=0.12in]\setlength\itemsep{0.25em}
    %%\item[$\mathbf{LOAD^{\DNSUD}_{AuthS}} (\mathbf{LOAD^{DNS}_{AuthS}})$:]
    \item[\bf{$\mathbf{LOAD^{DNSRU}_{AuthS}} (\mathbf{LOAD^{DNS}_{AuthS}})$}:]
    The number of requests to $AuthS$ for domain $D$ per second with (without) \DNSUD.
    \item [{\tt \bf{Update interval}} $I$:] The interval of time from the update 
    of domain $D$ record until $70\%$ of the resolvers in $R^D$ have queried $AuthS$ about $D$.
    If domain $D$ is highly popular, then most resolvers in $R^D$ query $AuthS$ at most $\Delta$  time after the update, to get a fresh copy of its record.
\end{description}

In the current DNS system, for popular domain $D$: 
\begin{center}
$LOAD_{AuthS}^{DNS} \approx \frac{R^D}{TTL}$ 
\end{center}
\ignore{
For example, the TTL of $amazon.com$ is $1_{min}$, and thus, every minute Amazon's $AuthS$ receives millions of requests. 
}
Note that if $D$ is evicted from the cache sooner, $LOAD_{AuthS}^{DNS}$ is even higher. 
%%$LOAD_{AuthS}^{\DNSUD}$ is lower than $LOAD_{AuthS}^{DNS}$ outside the {\tt update intervals}.
$LOAD_{AuthS}^{DNSRU}$ is lower than $LOAD_{AuthS}^{DNS}$ outside the {\tt update intervals}.
Each {\tt update interval} is slightly longer than $\Delta$ for highly popular domains.
For unpopular domains, the load on both methods is low.

\ignore{%%%%%%%%%%%%%%%%%%%%%%%%%%%%%%%%%%%%%%%%%%%%%
Consider a popular domain name $D$ (such as Amazon) whose DNS record is kept in authoritative server $AuthS$.

Amazon, Baidu, Readdit, Office.com, Ebay.com, 
Consider a domain name $D$ whose DNS record is kept in authoritative server $AuthS$.
Here, we develop a model for the load on authoritative servers (i.e., the number of requests to $AuthS$ per second) with and without the \DNSUD{}, assuming an ideal cache model. At a high level without \DNSUD{}, the load on authoritative servers with a small TTL value ($1-5$ minutes) is constantly high. While with \DNSUD{}, the load on $AuthS$ is sporadic most of the time (when $D$ is not immediately after an update) since most of the time, only new resolvers or those who recover from crashes query for $D$.
As explained in this paper, many Internet service owners decide to set a short TTL value for their domains, which produces excess traffic. In Section \ref{sec:ttl-unchanged-domains}, we measured the superfluous traffic from resolvers to authoritative servers. However, as mentioned in Section \ref{sec:freq-updates} more than 80\% of the examined domains have rarely updated their IP address.
Under an ideal cache model, the optimal wishful load on authoritative servers is obtained when the resolvers send requests only after an IP address change. 
}%%%%%%%%%%%%%%%%%%%%%%%%%%%%%%%%%%% 
%%%%%%%%%%%%%%%%%%%%%%%%%%%%%%%%%%%%%%%%%%%%%%%%%%%%%%%%%%%
%%%%%%%%%%%%%%%%%%%%%%%%%%%%%%%%%%%%%%%%%%%%%%%%%%%%%%%%%%%%%%%%

\subsection{\DNSUD{}: Feasibility and scalability}
\label{sec:solution-usage}

\ignore{%%%%%%%%%%%%%%%%%%%%%%%%%
While \DNSUD{} saves a significant amount of futile traffic between authoritative and resolver servers, it generates new traffic between the \UpdateDB{} and resolver servers. 
Here we show that \DNSUD{} can efficiently serve more than $520,000,000$ \emph{stable} domains ($=\#domains$) when the \UpdateDB{} sends UpDNS responses of size $< MaxSize$ (see Section \ref{sec:security}) and resolvers query the \UpdateDB{} every $\Delta = 1_{min}$. 
}%%%%%%%%%%%%%%%%%%%%%%%%%

Although \DNSUD{} effectively reduces futile traffic between authoritative and resolver servers, it introduces new traffic between the \UpdateDB{} and resolver servers. In this context, we demonstrate that \DNSUD{} can efficiently handle over $520,000,000$ \emph{stable} domains ($=\#domains$) by ensuring that the \UpdateDB{} sends UpDNS responses of size $< MaxSize$ (see Section \ref{sec:security}) and resolvers query the \UpdateDB{} every $\Delta = 1_{min}$.
The key observation is that since the domains serviced by \DNSUD{} are updated less than once in a few months on average (\emph{stable} domains), the total number of updates in $\Delta = 1_{min}$ is small enough to fit in an UpDNS response.

We set the following system variables:
\begin{description}%%[leftmargin=0.12in]\setlength\itemsep{0.25em}
    \item [UpDNS response size limit:] Let $\widehat{\#Mupdates}$ be the average number of records that fit into one UpDNS response, of size $MaxSize$. Following Section \ref{sec:security}, $MaxSize = 10,0000$, and each record takes $15$ bytes on average ($10$ bytes \cite{domain-length} for a domain name on average, $1$ for a subdomain flag, and $4$ for a timestamp). Therefore, $\widehat{\#Mupdates = 665}$.
    \item [$\mathbf{\Delta_{r}}$]: For every resolver $r$, \ $\Delta_{r} = \Delta = 1_{min}$.
\end{description}

Let $avg(freq_{update}) = \frac{1}{avg(T_{update})}$ be the average frequency of IP address updates for domains in the \UpdateDB{} per minute. The $avg(T_{update})$ for the \emph{stable} domains as given in Figure \ref{fig:domains-update} is $782,692 \ minutes > 17 \  months$.

Therefore, $\#domains$, the total number of domains that can be serviced by \DNSUD{}, is calculated as follows:
Multiplying the number of domain records that fit into a UpDNS response, sent every one minute, by the average number of minutes between IP address changes. Thus, we get:
$$\#domains = \frac{\widehat{\#Mupdates}}{\Delta \cdot avg(freq_{update})} = \frac{665}{1 \cdot avg(freq_{update})}$$ $$ = \frac{665}{\frac{1}{782,692}} = 520,490,000$$ 
This is larger than the estimated number of domains registered worldwide (some of which may not be relevant to \DNSUD) which is around 350 million domain names, according to \cite{num-domains}. \\

\textbf{Scalability:} \\
Here, we assess the number of requests that the \UpdateDB{} is expected to handle and its storage requirements, assuming all \emph{stable} domains worldwide use \DNSUD{} ($> 350M$ domains):
\begin{itemize}%%[leftmargin=0.12in]\setlength\itemsep{0.25em}
    \item The average number of updates the \UpdateDB{} receives from authoritatives per minute is $\widehat{\#Mupdates} = 665$.
    \item The average number of requests the \UpdateDB{} receives from resolvers every minute ($=\Delta$) bounded by the number of resolvers worldwide. 
    We estimate $100K$ as a plausible upper limit for this figure \cite{CiscoUmbrella}, which translates to roughly one resolver per $100K$ people. 
    Even a tenfold increase to $1M$ is still manageable.
    Notably, each request is processed by the server without the need for additional communication and involves a simple database query.
    \ignore{ %%%%%%%%
    Although the exact number of resolvers is unknown, \cite{root-load} indicates that approximately $10$ million requests are sent to each root DNS server every minute.
    Moreover, these sessions are easy for the \UpdateDB{} to handle as they consist of a single request and the efficient operations it performs for the response.
    Thus, if we integrate the \DNSUD{} service into the root system, we believe that \DNSUD{} would contribute a relatively minor additional load.
    } %%%%%%%%%%%
    \item \textbf{Storage size:} The total storage required for \UpdateDB{} is $12 MB$, since we store records for a maximum of a few hours.
    \ignore{$3 \cdot MaxHistory$ time (ranging from a few minutes up to three hour), see Section \ref{sec:update-process}.}
    Consequently, we only need to store about $12 MB$ which consists of tens of thousands of records ($665 \cdot 180 < 120K$), each of size $< 100B$.
\end{itemize}

\ignore{%%%%%%%%%%%%%%%%%%%%%%%%%%%%%%%%%%%%%%%%%%%%%%%%%%%
\begin{figure}[h]
    \centering
    \includegraphics[width=8.5cm]{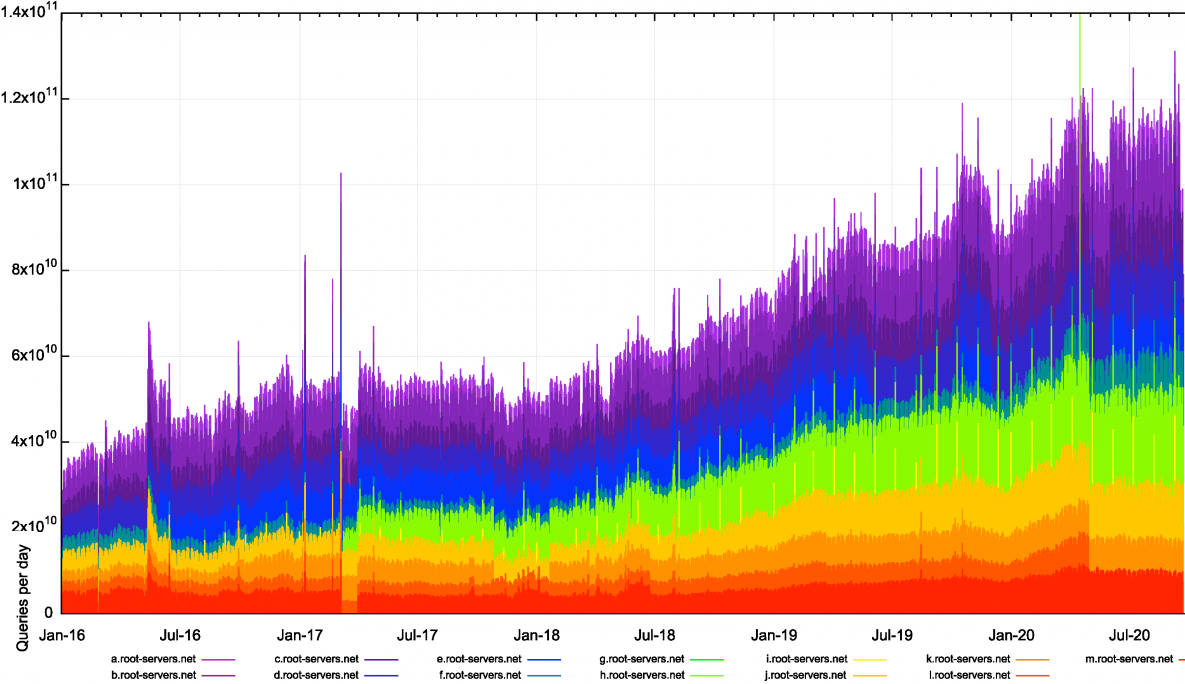}
    \caption{Traffic on root DNS servers per day}
    \label{fig:traffic-dns}
\end{figure}
}%%%%%%%%%%%%%%%%%%%%%%%%%%%%%%%%%%%%%%%%%%%%%%%%%%%%%%%%%%%

\ignore{%%%%%%%%%%%%%%
Section \ref{sec:freq-updates} examined $T_{update}$ for popular websites. 
Over half of the \emph{stable} domains have $T_{update} >$ 1 year, thus the average update time is more than 6 months. 

The total number of unique domain names that can use \DNSUD{} is thus:
$$\#domains = 520,490,000 \  | \  avg(freq_{update}) = \frac{1}{782,692}$$
}%%%%%%%%%%%%%%%%%%
\ignore{%%%%%%%%%%%%%%%%%%%%%%%%%%%%
\begin{itemize}\setlength\itemsep{0em}
    \item For an average update time of 6 months:
    $$\overline{freq_{update}^{minutes}} = \frac{1}{263,000}.\quad \#domains = 131,500,000 $$
    \item For an average update time of 1 year:
    $$\overline{freq_{update}^{minutes}} = \frac{1}{526,000}.\quad \#domains = 263,000,000 $$
\end{itemize}
\begin{figure}[h]
    \centering
    \includegraphics[width=8.3cm]{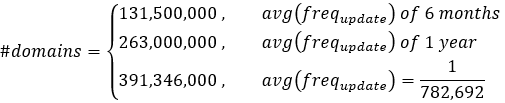}
    \label{fig:num-domains}
\end{figure}
}%%%%%%%%%%%%%%%%%%%%%%%%%%%%%%%%%%%

\ignore{%%%%%%%%%%%%%
However, if the $avg(freq_{update})$ is one year, we can support $263,000,000$ domains. 
}%%%%%%%%%%

\nsdionly{\textbf{Storage size:} The total storage required for \UpdateDB{} is at most $2.1MB$; 
The average size of a single record inside the write-only database is $15$ bytes (ten for domain, one for subdomains flag, and four for timestamp), and $8$ bytes for the hash table (four for timestamp, four for the pointer). As discussed in Section \ref{sec:update-process}, the maximum storage required at any point of time is the total number of records created in $3 \cdot MaxHistory$ time (in the order of a few minutes up to one hour). As noted above, about $500$ new records are created every minute, thus, the total storage required is at most $2.1MB$.}
%%%%%%%%%%%%%%%%%%%%%%%%%%%%%%%%%%%%%%%%%%%%%%%%%%%%%%%%%%%%
%%%%%%%%%%%%%%%%%%%%%%%%%%%%%
\subsection{Economic implications}
\label{sec:economic-implications}
%Due to space limitations the analysis of financial pros and cons of \DNSUD{} are eliminated from this submission.
%%Due to space limitations the analysis of \DNSUD{} financial pros and cons are removed from this submission.
%Due to space limitations, the DNSRU financial analysis has been removed (see \cite{arielThesis} for a full version).
%% back for Arxive \ignore{%%%%%%%%%%%%%%%%%%%%%%%%%%%%%%%  Space limmitations removed from submssion 
\ignore{%%%%%%%%%%%
Here we discuss possible economic implications that may result from using \DNSUD{}.\\
}%%%%%%%%%%
We consider the expenses and savings associated with the authoritative and recursive resolver servers.

\textbf{Authoritative servers:}
Two major costs for domain owners in the current DNS system that the \DNSUD{} may eliminate:
\begin{enumerate}%%[leftmargin=0.15in]\setlength\itemsep{0.25em}
%%%    \item \textbf{Futile Traffic:} The potential savings for all domain owners that use managed DNS services from reducing futile traffic are about hundreds of millions of dollars annually.
    \item \textbf{Futile Traffic:}
    We estimate that by reducing the number of requests to the authoritative server, domain owners using managed DNS services could potentially save hundreds of millions of dollars annually.
    This indicates the potential cost savings for managed services through \DNSUD{} adoption:
%%%    Our estimation is based on the following factors:
    \begin{enumerate}%%[leftmargin=0.15in]\setlength\itemsep{0.25em}
    \item \textbf{Per Request cost:} DNS queries sent to authoritative servers constitute a portion of the costs domain owners pay for managed DNS services. 
    The market rate per million requests is between $\$0.2$ to $\$0.4$ \cite{gcp, route53}. Here we assume an average of $\$0.3$ per million requests.  

    \item \textbf{Request Volume:}
    %%According to a public report of NS1 \cite{ns1}, NS1's collective authoritative servers consistently handle one million queries per second. 
    %%Since NS1 is about $4\%$ of the managed DNS market, we estimate that the total market services around $25M$ queries per second. 
    According to a report from NS1 \cite{ns1}, their authoritative servers handle one million queries per second. Considering NS1's $4\%$ market share, we estimate the total market services approximately 25 million queries per second.

    \item \textbf{Potential savings with \DNSUD:} As discussed in Section \ref{sec:ttl-unchanged-domains}, \DNSUD{} nearly eliminate all futile traffic of DNS queries that return a NOERROR response for \emph{stable} domains. 
    According to the data, the queries that were resolved with a NOERROR response amount to about $87\%$ \cite{ns1}. Meanwhile, the proportion of \emph{stable} domains exceeds $80\%$, (Section \ref{sec:freq-updates}).     
    Hence, \DNSUD{} would save about $70\%$ of the cost.
    \end{enumerate}

    From points (a), (b), (c) and considering the number of seconds in a year, the potential annual savings is: \\
    $\$0.3 \cdot 25 \cdot 70\% \cdot 31,556,926 \approx \$165,000,000$.

    Notice that the managed DNS services pricing model is more complicated in actuality.
    The \$$165,000,000$ is a rough estimation.
    Furthermore, a considerable number of domain owners use private (unmanaged) DNS servers, that is, the total global savings for domain owners could be substantially higher. 

    \ignore{%%%%%%%%%%%%%%%%%%%%%%%%%%%%%%%%%%%%%%%%%%%
    In section \ref{sec:ttl-unchanged-domains}, we discussed the potential traffic savings by authoritative servers due to the \DNSUD{} adoption.
    With \DNSUD{} authoritative server owners could save hundreds of millions of dollars every year.
    To estimate this cost, we rely on a public report from NS1 regarding their managed DNS services \cite{ns1}. 
    According to the report, NS1's authoritative servers consistently handle one million queries per second while charging their customers $\$5$ for every million queries.
    Consequently, NS1 bills all their clients approximately $\$150M$ per year solely for the DNS queries directed to their authoritative servers. 
    With \DNSUD{}, a significant portion of this expense could be mitigated since more than $80\%$ of domains are \emph{stable}.
    }%%%%%%%%%%%%%%%%%%%%%%%%%%%%%%%%%%%%%%%%%%%%%%%%%%%%%
    \item \textbf{Unavailability costs:}
    As discussed in the Introduction, the high-availability aspect of \DNSUD{} saves prolonged service unavailability due to misconfigurations in authoritative servers, which may result in revenue losses.
\end{enumerate}
\ignore{%%%%%%%%%%%%%%%%%%%%%%%%%%%%%%%%%%%%%%%%%%%%%%%%%%%
\begin{figure}[h]
    \centering
    \includegraphics[width=8.5cm]{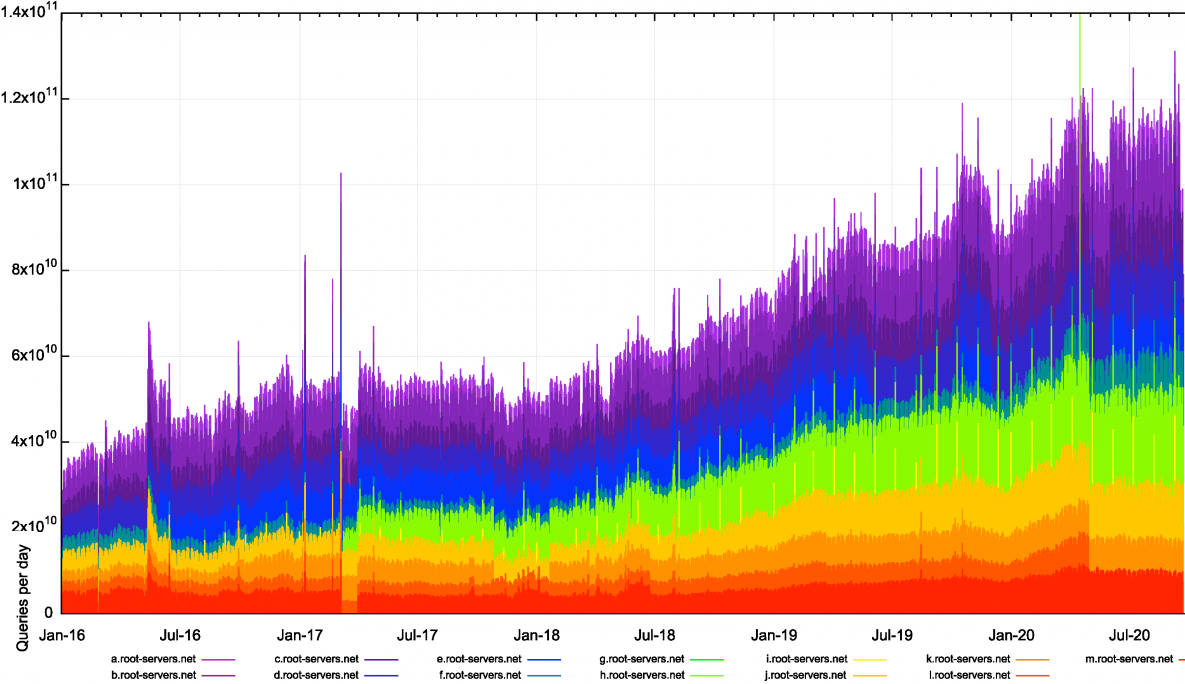}
    \caption{Traffic on root DNS servers per day}
    \label{fig:traffic-dns}
\end{figure}
}%%%%%%%%%%%%%%%%%%%%%%%%%%%%%%%%%%%%%%%%%%%%%%%%%%%%%%%%%%%
While \DNSUD{} eliminates these two losses, it also provides an additional gain and introduces a new charge on the authoritative servers for using the \DNSUD{} service:
\begin{itemize}%%[leftmargin=0.12in]\setlength\itemsep{0.25em} 
    \item (Potential 3rd Gain) \textbf{Transitioning authoritative servers to scalable Services:} Increasing the TTL values associated with domain names in \DNSUD{} changes the traffic patterns on authoritative servers. After updating a record's IP address, there is a peak followed by a prolonged period of low load, as illustrated in Section \ref{sec:load-dns} and Fig. \ref{fig:load}. Consequently, authoritative servers might benefit from transitioning to scalable cloud services and consolidating multiple zone files into a single server, further minimizing their costs.
    \item \textbf{Cost of using \DNSUD:} Maintaining \DNSUD{} might be a costly operation \cite{eliminating-dns} and this cost could be charged back to domain owners. The analysis of the economic trade-off between this cost and the three aforementioned gains is relegated to the full version.
\end{itemize}

\textbf{Recursive resolvers:}
The economic motivation for resolvers might not be as potent as it is for authoritative servers. Nonetheless, resolvers still benefit from reducing the futile traffic. 
One strategy to enhance their incentive to adopt \DNSUD{} could be to share the revenues earned from payments made by authoritative servers (domain owners) for using \DNSUD.
%%with them the revenues obtained from authoritative (domain owners) for using \DNSUD{}.

\ignore{%%%%%%%%%%%%%%%%%%%
\noindent
\textbf{Economic implications on resolvers:}\\
\DNSUD{} could raise the load on resolvers due to the following implications:
\begin{itemize}[leftmargin=0.12in]\setlength\itemsep{0.25em}
        \item \textbf{Increasing resolver's cache:} \DNSUD{} usage creates motivation to increase the resolver's cache size to prevent the eviction of DNS records from their local cache before the new larger TTL expires.
        \item \textbf{Reducing the traditional TTLs:} After \DNSUD{} usage, longer "\DNSUD-TTLs" are sent to the resolvers, see Section \ref{sec:new-ttl}. In addition, authoritative servers may reduce the traditional TTLs since the short TTLs sent to clients would improve the recovery time while increasing the load on resolvers instead of on authoritative servers.
\end{itemize}

}%%%%%%%%%%%%%%%%%%%%%%%%%%%%%%%%%%%%%%%%%%%%%%%%%%%%%%%%%%%%%%%%%%%%%%%%%%
%%%%%%%%%%%%%%%%%%%%%%%%%%%%%%%%%%%%%%%%%%%%%%%%%%%%%%%%%%%%%%%%%%%%%%%

%%%% back for arxive }%%%%%%%%%%%%%%%%%%%%%%%%%%%%%%%  Space limmitations removed from submssion 

%%% Local Variables:
%%% mode: latex
%%% TeX-master: "cwc"
%%% End:

%% file: security.tex
%\newpage \newpage
\section{Security}
\label{sec:security}
%%%%%%%%%%%%%%%%%%%%%%%%%%%%
The \DNSUD{} implementation must ensure that it does not open the door to any attack, such as a DDoS, DNS poisoning, on elements of the DNS system. 
The threat model includes $4$ potential threats which we address in this section:
\begin{enumerate}%%[leftmargin=0.15in]\setlength\itemsep{0.25em}
    \item DoS/DDoS on the \DNSUD{} that impairs its availability - This becomes notably severe since many domains are expected to use much larger TTL values with this service, see Section \ref{sec:increase-ttl}, and rely on it to update their IP address.
    \item \UpdateDB{} as a reflector - The \UpdateDB{} may be used as a reflector \cite{reflection, reflection2} in an attack on resolvers or any target on the Internet. In this attack, many UpDNS requests are sent to the \UpdateDB, with a source IP address spoofed to that of a target victim (resolver or any other victim); the victim then receives a flood of UpDNS responses from the \UpdateDB.
    \item
    Flooding Attack: An attacker floods \UpdateDB{} with InsertDomain requests, causing resolvers to receive large UpDNS responses filled with numerous domain records, thereby burdening them with substantial loads.
    \ignore{%%%%%%
    Flooding attack - An attacker could frequently send InsertDomain requests to \UpdateDB{}. When these requests are sent repeatedly, they can flood the \UpdateDB{} with numerous domain record insertions. This, in turn, prompts \UpdateDB{} to send large UpDNS responses to the resolvers.
    }%%%%%%
    \item Integrity of records stored by \UpdateDB{} - An attacker could manipulate the \UpdateDB{} to send an incorrect list of domains to the resolvers by either: inserting invalid domain names (which have not been updated) to the \UpdateDB{} or spoofing the \DNSUD{} service to send an invalid UpDNS response to the resolvers. This could lead to resolvers eliminating records from their cache that are still valid which may lead to three types of attack:
    \begin{enumerate}%%[leftmargin=0.15in]
    \item DoS on DNS authoritative servers - If the attack deletes many records from many resolvers' caches, it increases the load on the corresponding authoritative servers.
    \item Poor client experience - This attack might also affect the latency experienced by resolver's clients \cite{moura2019cache}.
    \item Load on the \UpdateDB{} - This attack could result in plenty of records inserted into \UpdateDB{}, potentially causing \UpdateDB{} to send many records to resolvers.
    \end{enumerate}
\end{enumerate}
Notice that DNS record poisoning is impossible in \DNSUD{}.

We propose these mitigations to deal with the above threats: 
\begin{description}%%[leftmargin=0.12in]\setlength\itemsep{0.25em}

  \item [Static IP address (threat $1$):] Since \DNSUD{} aims to provide high availability to DNS services, we avoid relying on DNS mappings to access it. As such, we recommend integrating \DNSUD{} into the root system with a static IP address for the \UpdateDB{} to prevent \DNSUD{} from becoming a new SPOF.
  
  \item [Reflection attack (threat $2$)] can be prevented by various anti-spoofing DDoS mitigation techniques \cite{dos-tech, dos-tech2}. Here, we propose that the \UpdateDB{} detects the anomalous activity of frequent, repeated requests originating from the same source IP address. In such instances, the \UpdateDB{} service responds with a TQ bit  (similar to the TC bit in the standard DNS protocol), indicating that for a certain period, it accepts requests from this IP address only over QUIC.
  
  \item [Bounding UpDNS response size (threat $2$):] 
  A recovering \\ resolver or a malicious resolver could query for old records stored in the \UpdateDB{}, thereby forcing it to dispatch large UpDNS responses. To mitigate this reflection attack, we limit the size of UpDNS responses to $10,000$ bytes, fitting into a small number of packets.
  \ignore{, as described in Section \ref{sec:update-process}.}
  \ignore{%%%%%%
  A recovering resolver or a malicious resolver could query for old records stored in the \UpdateDB{}, thereby forcing it to dispatch large UpDNS responses. 
  To prevent this large-scale amplification attack we bound the size of UpDNS response messages. 
  Let $MaxSize$ be the maximum size of a response.
  A resolver that queries the \UpdateDB{} with an old timestamp, $t_{old}$, such that the \UpdateDB{} has $M (> MaxSize)$ records in the database from $t_{old}$ to the current time, undergoes a series of UpDNS requests and responses until it obtains all $M$ records in the database from $t_{old}$ up to the current time.
  Each request's timestamp is the maximum of the previous response's timestamp in the sequence.
  We set $MaxSize$ to $10,000$ bytes, fitting into a small number of packets ($<7$).}
  
  \item [Flooding attack (threat $3$)] 
  %can be prevented by authenticity techniques.
  To thwart this threat, \UpdateDB{} \\ should detect unusual patterns, such as repeated InsertDomain requests from specific IP addresses or targeting similar domains. Then enforce validation of InsertDomain requests from suspicious IP addresses for a certain period. Options include; server registration, request rate limits for new servers, or CAPTCHAs.
  \ignore{%%%%
  We suggest that the \UpdateDB{} detects anomalous activity characterized by frequent InsertDomain requests, either originating from a small set of IP addresses or targeting similar domain names.
  As a mitigation measure, \UpdateDB{} should require authoritative servers to validate their InsertDomain requests for a certain period, using methods like requiring authoritative servers to register with the \UpdateDB{} service, throttling the InsertDomain request rate from new authoritative servers, or suspicious IP addresses, or Captchas. 
  }%%%%
  \nsdionly{ \item [Flooding attack ] 
  can be prevented by authenticity techniques. Here, we suggest that the \UpdateDB{} detects anomalous activity characterized by frequent InsertDomain requests, either from the same small set of IP addresses or for similar domain names. When these requests are sent repeatedly by an attacker, they can flood the \UpdateDB{} with numerous domain record insertions. 
  This, in turn, prompts \UpdateDB{} to send large UpDNS responses to the resolvers. 
  As a mitigation the \UpdateDB{} should require authoritative servers to validate their InsertDomain requests for a certain period, using methods like Captchas, requiring authoritative servers to register to the \UpdateDB{} service, throttling InsertDomain request rate from new authoritative servers, or suspicious IP addresses. 
  }

  \item  [Authenticity of]\textbf{messages (threat $4$):} 
  \begin{enumerate}%%[leftmargin=0.15in]\setlength\itemsep{0em}
    \item We suggest two methods to ensure the authenticity of the UpDNS responses, first using a nonce (a random number issued in an authentication protocol) generated by the resolver and second, using QUIC \cite{rfc9000} for the communication between the \UpdateDB{} and the resolvers to prevent the MITM attack. For this to work, the \UpdateDB{} must register with a certificate authority to acquire an authentic certificate.
    In the former, the \UpdateDB{} includes the timestamp and the nonce from the corresponding request in the UpDNS response, effectively preventing spoofed or replay attack \cite{playback} on the resolver.
    \item To ensure the authenticity of the InsertDomain requests, we apply two methods, first including a timestamp in the InsertDomain request and second, signing the request with DNSSEC \cite{dnssec-view2, rfc4035}.
    %\cite{dnssec-view, dnssec-view2, rfc4035}. 
    Lastly, to enhance reliability, this communication occurs over TCP. In the former, the \UpdateDB{} validates that the timestamp corresponds to a short time before the current time, thus disabling replayed InsertDomain requests.
  \end{enumerate}
  
  \item [Hardening the \DNSUD{}: (threats $1, 4$)] In order to minimize the attack surface on the \UpdateDB{} we employ security best practices, such as installing security patches, strong authentication, secure code development,  DDoS protection solutions, and secure management practices. \cite{attack1, attack2, attack3}.
  \item [Switching \DNSUD{} to DNS (threat $1$):] One potential strategy to mitigate DoS attacks on the \DNSUD{} service could be to revert it to the standard DNS system. Specifically, if the UpdateDB becomes unavailable for a resolver, the resolver can ignore the \DNSUD-TTL and use the traditional one. \ignore{While this idea preserves the current availability of DNS, it adds complexity for resolvers that must maintain two TTLs for every cache record.}
\end{description}

%%%%%%%%%%%%%%%%%%%%%%%%%%%%%%%%%%%%%%%%%%%%%%%%%%%%%%
\nsdionly{\subsection{DNS unicast routing support} 
\label{sec:unicast}
\DNSUD{} allows Internet services to use unicast DNS services \cite{unicast}, i.e., provide different IP addresses for different resolvers.
When the IP address of domain $D$ changes, every resolver deletes $D$ from its cache. Subsequently if necessary it resolves $D$ to obtain a fresh value in the standard DNS system.}
\ignore{%%%%%%%%%%%%%%%%%%%%%%
For resolvers using \DNSUD{} it is possible to ensure a constant update time for a domain name record that has been updated after the authoritative server updates the \DNSUD{}. Thus, \DNSUD{} allows Internet service owners to change their TTLs: if the duration for total recovery (i.e., all clients receive a correct resolution for a domain name after updating its record) is constant and short (e.g., $\Delta=1_{min}$), they can increase their TTLs indefinitely, reducing the traffic to their authoritative server and the economic costs as a result. In Section \ref{sec:freq-updates}, we found that more than 99\% of the examined domain names choose TTL above $1_{min}$, so for these domains, there is no compromise on their duration for total recovery.
}%%%%%%%%%%%%%%%%%%%%%%%%%%%%
%%%%%%%%%%%%%%%%%%%%%%%%%%%%%

%% file: drill-down.tex
\section{\DNSUD{} drill-down}
\label{sec:drill-down}
%%%%%%%%%%%%%%%%%%%%%%%%%%%%%%%%%%%%%%%%%%%%%%%%%%%
%%%%%%%%%%%\subsection{Garbage collection of \UpdateDB{} records}
\subsection{\UpdateDB{} structure and messages}
\label{sec:update-process}

In this section, we explain the database structure and the process that the \DNSUD{} service performs when it receives a request.
%%%%%%%%%%%%%%%%%%%%%%%%%%%%%%%%%%%%%%%%%%%%%%%%%%%%%
Each record in the \UpdateDB{} holds a recently updated domain name, with the following fields:
\begin{itemize}%%leftmargin=0.12in]\setlength\itemsep{0.25em}
    \item Domain - The domain whose record has been recently updated (it does not have to be a parent domain)
    \item Subdomains - A flag indicating whether the domain's subdomains should also be removed from the resolvers' cache.
    \item Timestamp - The time at which the record was inserted into the \UpdateDB.
\end{itemize}

As in the DNS system the UpDNS request and response are sent over UDP to enable quick and efficient communication. For reliability, the InsertDomain message is sent over TCP.
In Section \ref{sec:security}, for security reasons, QUIC is suggested for the UpDNS request and response, and the use of DNSSEC signatures for the InsertDomain request is recommended. The format of these messages is as follows:
\begin{enumerate}%%[leftmargin=0.15in]\setlength\itemsep{0.25em}
    \item The InsertDomain request includes the domain name, its subdomain flag, and a timestamp to mitigate replay attacks, as discussed in Section \ref{sec:security}.
    \item The UpDNS request contains the maximum timestamp received in the previous response from the \UpdateDB{} and a nonce for security reasons (see Section \ref{sec:security}).
    \item The UpDNS response contains a list of domain names that were inserted into the \UpdateDB{} after the timestamp received in the request, the request's timestamp, and nonce.
\end{enumerate}

\textbf{\UpdateDB{} responses:}\\
Upon receiving an InsertDomain request from an authoritative server, the \UpdateDB{} first verifies the request as detailed in the security section, \S\ref{sec:security}, and then inserts a new record into the \UpdateDB{} with the following fields: the received domain, the subdomains flag, and the current time.

Each domain in \UpdateDB{} is associated with a unique timestamp indicating the last time the domain name was inserted into the \UpdateDB.
Upon receiving an UpDNS request from a recursive resolver containing the timestamp associated with the latest update this resolver has received, the \UpdateDB{} finds all domains that have been inserted into the \UpdateDB{} after the request's timestamp and sends them back to the resolver with the request's timestamp and a nonce.

To prevent large-scale amplification reflection attacks and other security reasons we bound the size of UpDNS response messages.
Let $MaxSize$ be the maximum size of a response.
A resolver that queries the \UpdateDB{} with an old timestamp, $t_{old}$, such that the \UpdateDB{} has $M (> MaxSize)$ records in the database from $t_{old}$ to the current time, undergoes a series of UpDNS requests and responses until it obtains all $M$ records in the database from $t_{old}$ up to the current time.
The timestamp of each request is the maximum timestamp of the previous response in the sequence.
We set $MaxSize$ to $10,000$ bytes, fitting into a small number of packets. 
In Section \ref{sec:security}, we recommend to use QUIC for this exchange. 
If pure UDP is used and the UpDNS response is larger than the MTU ($1500$ bytes), the \UpdateDB{} replies with a TQ bit (similar to the TC bit in the standard DNS protocol), requesting the resolver to resend the UpDNS request over QUIC.

\textbf{Resolver operations:}\\
Every $\Delta$ time units, the resolver issues an UpDNS request.
Upon receiving the UpDNS response, which includes a list of domains and their respective timestamps, the resolver deletes these domains from its cache, and stores the maximum timestamp for use in the subsequent UpDNS request.
If the Subdomains flag is set for a domain, then all of its subdomains are also removed.
The added load on the resolvers while processing the UpDNS response is negligible since the heaviest operation in this is to access (and delete) elements from its local cache for each domain in the list, a process which requires on the order of 100 machine cycles per removal. The number of these records is limited to $665$, as discussed in Section \ref{sec:solution-usage}.

\begin{figure}[h]
    \centering
    \includegraphics[width=8cm]{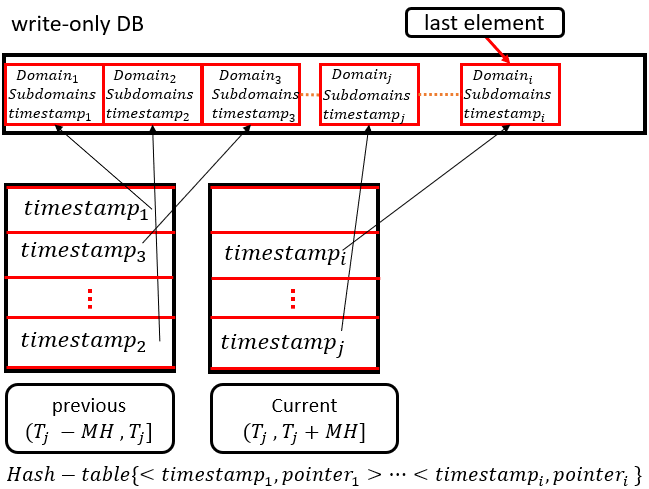}
    \caption{\UpdateDB{} data structure schema. $MH = MaxHistory$}
    \label{fig:data-structure}
\end{figure}

%%%%%%%%%%%%%%%%%%%%%%%%%%%%%%%%%%%%%%%%%%%%%%%%%%%%%
%%%%%%%%%%%%%%%%%%%%%%%%%%%%%%%%%%%%%%%%%%%%%%%%%%%%%%
%%%%%%%%%%%%%%%%%%%%%%%%%%%%%%%%%%%%%%%%%%%%%%%%%%%%%%
\textbf{\UpdateDB{} two data structures (see Figure \ref{fig:data-structure}):}
\begin{itemize}%%[leftmargin=0.12in]\setlength\itemsep{0.25em}
    \item A write-only DB - Each element in this database contains a domain name, subdomains flag, and timestamp. A pointer to the last element in the DB is maintained. The write-only database is kept in a linear array. Thus it is easy to sequentially read elements from a given entry to the end.
    \item A hash table keyed by a timestamp - For each element in the write-only DB, a corresponding element is generated in the hash table, keyed by the its timestamp, and valued with a pointer to the corresponding element in the DB. The hash table garbage collection is detailed in the sequel.
\end{itemize}
%%%%%%%%%%%%%%%%%%%%%%%%%%%%%%%%%%%%%%%%%%%%%%%%%%%%%%
Records inserted into the \UpdateDB{} need to be garbage collected at some point in time. We define a system parameter, $MaxHistory$, which is the minimum duration of time each record must be kept in the database.
%%$MaxHistory$ is the maximum time that \UpdateDB{} wants to store domain names after they have been inserted into the "write only" DB.
Essentially, $MaxHistory$ is the maximum value between the following two:
\begin{enumerate}%%[leftmargin=0.15in]\setlength\itemsep{0.25em}
    \item The maximum $\Delta$ that any resolver may use. 
    \item The oldest record that a resolver recovering from a crash or a newly joining resolver might need (see Section \ref{sec:joining-process}).
\end{enumerate}

To implement the database garbage collection the time is divided into an infinite sequence of epochs of length $MaxHistory$ $(MH)$, i.e.,
$\{[0, MH], (MH, 2MH],\ldots , ((i-1)MH, iMH]\},\ldots\}$. 
We maintain two hash tables that cover two consecutive epochs. 
The {\tt current hash table} covers the current time epoch, and is where new records are inserted, while the {\tt previous hash table} contains the elements of the previous epoch.

\ignore{%%%%%%%%%%%
interval is for the current half an hour and the previous is for the last half an hour, i.e., if the current time is 11:47, the {\tt previous interval} contains entries for elements that have been inserted in the $(11:00, 11:30)$ interval. In comparison, the {\tt current interval} contains entries for elements that have been inserted in the $(11:30, 12:00)$ interval. 
}%%%%%%%%%%%
$MaxHistory$ is a design parameter, which we believe should be about a few minutes up to one hour. 
\nsdionly{
Note that if caches become non-volatile (NVM memory) in the future, then the value of $MaxHistory$ should be reconsidered because recovering resolvers may recover with an outdated (NVM) cache. Thus, either $MaxHistory$ should be in the order of the maximum failure time, or recovery from a long failure will require clearing the entire cache, see Section \ref{sec:joining-process}.
}

%%%%%%%%%%%%%%%%%%%%%%%%%%%%%%%%%%%%%%%%%%%%%%%%%%%%%
\textbf{\UpdateDB{} operations:}\\
{\bf Finding domain names that come after a timestamp $\textbf{{t}}$:} 
In response to an UpDNS request with timestamp $t$, the \UpdateDB{} returns the list of domains that have been inserted after time $t$, as follows:
\begin{enumerate}%%[leftmargin=0.15in]\setlength\itemsep{0.25em}
    \item Using the hash tables, find $e_t$, the last element in the write-only DB that is associated with time $t$.
    \item Return all the elements that follow $e_t$ in the write-only DB to the head of the series (the last element).
\end{enumerate}
Let $n$ represent the number of such elements.\\
The {\em Time complexity} of this operation is $O(1)$ to find the element in the write-only DB, and $O(n)$ to return the list of elements that follow $e_t$. According to Section \ref{sec:solution-usage}, $n$ is less than $700$ on average.

\noindent
{\bf Inserting a new domain name:} 
Upon receiving an InsertDomain request($Domain, Subdomains Flag$) from an authorized server the \UpdateDB{} takes the following steps:
\begin{enumerate}%%[leftmargin=0.15in]\setlength\itemsep{0.25em}
    \item Append a new entry at the end of the write-only DB with \texttt{Domain} = $Domain$, \texttt{Subdomains} = $Subdomains Flag$\ \ and \texttt{timestamp} $ts$ = current time.
    \item Update the \texttt{last element} pointer to $p$, the pointer to the newly inserted record.
    \item Insert a new entry into the hash table with a \texttt{value} equal to $p$, and a \texttt{key} equal to $ts$.
\end{enumerate}

The {\em Time complexity} of this operation is O(1) to add a new element at the end of the write-only DB and O(1) to update the hash table.

\noindent
{\bf Deleting old records from the \UpdateDB{}:} 
As explained earlier, \UpdateDB{} can delete domain names that have been inserted more than $MaxHistory$ time ago. When the current epoch reaches its end, \UpdateDB{} performs the following:
\begin{enumerate}%%[leftmargin=0.15in]\setlength\itemsep{0.25em}
    \item \UpdateDB{} records a pointer, $ptr$, to an arbitrary record in the write-only database that corresponds to a timestamp in the {\tt previous hash table}. This record is used in the sequel (step (4) to garbage collect all the records that correspond to the {\tt previous interval}.
    \item Sets the {\tt current hash table} as the {\tt previous hash table}.
    \item Allocates a new hash table for the current time interval. 
    \item Starts a process, that concurrently with other operations, lazily deletes all the records before and after the record pointed by $ptr$ that belong to the epoch preceding the currently {\tt previous interval}.
\end{enumerate}

Note that the \UpdateDB{} starts to delete each element at most $2 \cdot MaxHistory$ time after it was inserted. In our implementation, we should delete elements from the write-only DB, while the system garbage collection will free the hash table after it's no longer in use. We assume that the deletion process completes before the next deletion process starts (after $MaxHistory$ time). Let $k_p$ be the number of elements in the {\tt previous interval}.

The {\em Time complexity} of this operation is $O(1)$ for steps (1) through (3), and $O(k_p)$ time to delete the records that belonged to the interval that is being deleted. We anticipate it takes more time for the system garbage collection to free all the records that have been deleted. However, step (4) is executed lazily and concurrently with the system operations.

%%%%%%%%%%%%%%%%%%%%%%%%%%%%%%%%%%%%%%%%%%%%%%%%%%%%%%
%%%%%%%%%%%%%%%%%%%%%%%%%%%%%%%%%%%%%%%%%%%%%%%%%%%%%%

\ignore{%%%%%%%%%%%%%%%%%%%%%%%%%%%%%%%%%%%%%%%%%%%%%%%%%%%%%
\subsection{\DNSUD{} messages and communications}
\label{sec:message-structure}
Choosing the suitable transport layer communication protocol between the \UpdateDB{} and the recursive resolvers or authoritative servers is essential for the efficiency and security of the system.
To enable quick, fast, and efficient communication between the resolvers and the \UpdateDB{}, we suggest sending the messages between resolvers and \UpdateDB{} over UDP as in the standard DNS system \cite{dns-udp1, dns-udp2}.
The communication between the authoritative servers and the \DNSUD{} should be reliable. In Section \ref{sec:freq-updates}, it is argued that \DNSUD{} is used mostly by "stable domains" - domains whose IP address is updated less than once a month. Therefore, TCP is used for this communication. In Section \ref{sec:security}, we discuss options to secure these channels in order to mitigate various attacks.
}%%%%%%%%%%%%%%%%%%%%%%%%%%%%%%%%%%%%%%%%%%%%%%%%%%%%%%%%%%%%%%%%%
%%%%%%%%%%%%%%%%%%%%%%%%%%%%%

\subsection{Adding a new resolver to the service}
\label{sec:joining-process}
A new resolver that joins the \DNSUD{} service at time $T$ should perform the following actions: 
\begin{enumerate}%%[leftmargin=0.15in]\setlength\itemsep{0.25em}
    \item Flush its local cache.
    \item Set its local $timestamp$ to $T$, i.e., in the first UpDNS request, the $timestamp = T$.
    \item Set a timer to send its first UpDNS request in $\Delta$ time units.
\end{enumerate}
In this way, a cache miss occurs when a client queries the resolver to resolve a domain name. Subsequently, the resolver resolves the query in the standard DNS system to receive a fresh DNS response. When the timer expires, it queries the \UpdateDB{} for any record that may have been updated in the last $\Delta$ time units. From this point, the resolver operates as described in the previous sections.
The joining process is also performed by recursive resolvers that recover from a crash or lose their Internet connection for more than $MaxHistory$ time. Resolvers that lose their network connection for less than $MaxHistory$ time operate as usual, see Section \ref{sec:update-process}.   

\ignore{%%%%%%%%%%%%%%%%%%
after recovering from a crash. When should they perform the joining process?
On the one hand, we do not want the \DNSUD{} service to handle UpDNS requests with old timestamps. Thus, from the narrow perspective of the \DNSUD{} service, we prefer a short duration (several minutes) because the joining process initiates the timestamp to the current time. On the other hand, the joining process contains flushing a resolver's local cache. It can increase the resolver and authoritative server load and create excessive traffic.
} %%%%%%%%%%%%%%%%%%

%%%%%%%%%%%%%%%%%%%%%%%%%%%%%%%%%%%%%%%%%%%%%%%%%%%%%%
\subsection{DNS routing features support} 
\label{sec:unicast}
%In the evolution of DNS, crucial networking features were developed to ensure efficient and fast DNS resolutions.
Since \DNSUD{} only removes domain records from a resolver's cache and does not affect the actual resolution process, the traditional DNS system continues to use any routing feature (anycast, unicast, round-robin \cite{unicast, roundrobin}, etc.) as before. 
Moreover, anycast can be used in the implementation of \DNSUD{} to ensure its efficient and fast response time.
\ignore{%%%%%%%%%%%%%%%%%%%%%%
For resolvers using \DNSUD{} it is possible to ensure a constant update time for a domain name record that has been updated after the authoritative server updates the \DNSUD{}. Thus, \DNSUD{} allows Internet service owners to change their TTLs: if the duration for total recovery (i.e., all clients receive a correct resolution for a domain name after updating its record) is constant and short (e.g., $\Delta=1_{min}$), they can increase their TTLs indefinitely, reducing the traffic to their authoritative server and the economic costs as a result. In Section \ref{sec:freq-updates}, we found that more than 99\% of the examined domain names choose TTL above $1_{min}$, so for these domains, there is no compromise on their duration for total recovery.
}%%%%%%%%%%%%%%%%%%%%%%%%%%%%
%%%%%%%%%%%%%%%%%%%%%%%%%%%%%

\subsection{\DNSUD{}: Bootstrapping the adoption}
\label{sec:market}
%%%%%%%%%%%%%%%%%%%%%%%%%%%%%%%%%%%
Common resolvers lack motivation to adopt the \DNSUD{} method until a significant number of authoritative servers do so. 
However, major DNS public providers like Cloudflare, Google, and Amazon (see \cite{auth-market}) have two key reasons to embrace the method:
\begin{enumerate}
    \item They dominate a considerable portion (close to $50\%$) of the open resolvers market (as shown in Table \ref{table:resolvers-market}), which means they can realize immediate savings. 
    Consequently, if Google and Cloudflare adopt the service, they would cover about $50\%$ of the DNS resolutions and create a compelling incentive for both resolvers and authoritative servers to follow and adopt the \DNSUD.
    \item As these prominent DNS name server providers adopt the method, it incentivizes other resolvers to do the same, resulting in additional savings for both parties. 
\end{enumerate}
Supporting the above points is the recent introduction of the "cache flushing service" by Cloudflare and Google, see Section \ref{sec:background}.

\ignore{%%%%%%%%%%rotating table
\begin{table}[h]
\centering
\begin{tabular}{|c|c|}
\hline
\multicolumn{1}{|c|}{\textbf{DNS Provider}} & \textbf{Share (in 2019)} \\ \hline
Google                                       & 35.94\%               \\
Cloudflare                                       & 13.80\%              \\
Quad9                                       & 0.78\%                 \\
Yandex                                       & 0.09\%                 \\
OpenDNS                                       & 0.03\%                 \\
\hline
\end{tabular}
\caption{Total public resolvers market share from \cite{resolver-market, reslover-market2}}
\label{table:resolvers-market}
\end{table}
%%%%%%%%%%%%%%%%%%%%%%%%%%
}%%%%%%%%%%rotating table

\begin{table}[h]
\centering
\begin{tabular}{|c|c|c|c|c|c|}
\hline
\multicolumn{1}{|c|}{\textbf{DNS Provider}} & \textbf{Google} & \textbf{Cloudflare} & \textbf{Quad9} & \textbf{Yandex} & \textbf{OpenDNS} \\ \hline
\multicolumn{1}{|c|}{\textbf{Share (in 2019)} } &  35.94\%  &  13.80\% & 0.78\% & 0.09\% &  0.03\% \\ \hline
\end{tabular}
\caption{Total public resolvers market share from \cite{resolver-market, reslover-market2}}
\label{table:resolvers-market}
\end{table}

%%%%%%%%%%%%%%%%%%%%%%%%%%%%%%%%%%
\ignore{%%%%%%%%%%
The adoption of \DNSUD{} is closely linked to major players in DNS outsourcing, particularly Google and Cloudflare, as they dominate the open resolver market (Table \ref{table:resolvers-market}). The economic incentives for authoritative servers to adopt \DNSUD{} are compelling, especially considering the influence of Google and Cloudflare in the authoritative servers market \cite{auth-market}. Integration of \DNSUD{} by these giants could lead to approximately $50\%$ of the potential savings for authoritative servers using this technology.
}%%%%%%%%%%

\nsdionly{%%%%%%%%%%%%%%%%%%%%%%%%%%%%%%%%%%
\begin{table}[h]
\centering
\begin{tabular}{|c|c|}
\hline
\multicolumn{1}{|c|}{\textbf{DNS Provider}} & \textbf{Share (in 2019)} \\ \hline
Google                                       & 35.94\%               \\
Cloudflare                                       & 13.80\%              \\
Quad9                                       & 0.78\%                 \\
Yandex                                       & 0.09\%                 \\
OpenDNS                                       & 0.03\%                 \\
\hline
\end{tabular}
\caption{Total public resolvers market share from \cite{resolver-market, reslover-market2}}
\label{table:resolvers-market}
\end{table}
\ignore{%%%%%%%%%%%%%%%%%%%%%%
Table \ref{table:resolvers-market} shows that if Google and Cloudflare (which already have DNS flushing service, see Section \ref{sec:background}) use the \DNSUD{} service, they cover about $50\%$ of the DNS resolutions.}
}%%%%%%%%%%%%%%%%%%%%%%%%%%%
%%%%%%%%%%%%%%%%%%%%%%%%%%

\nsdionly{%%%%%%%%%%%%%%%%%%%%%%%%%%%%%%%%%%%%%%%%%%%%%%%%%%%%%%%%%%%%%%
\begin{table}[h]
\centering
\begin{tabular}{|c|c|}
\hline
\multicolumn{1}{|c|}{\textbf{CDN Provider}} & \textbf{Share} \\ \hline
Cloudflare                                       & 19.1\%               \\
Fastly                                       & 1.7\%              \\
Amazon                                       & 1.4\%                 \\
Sucuri                                       & 0.5\%                 \\
Akamai                                       & 0.4\%                 \\
\hline
\textbf{Others}                                       & \textbf{76.1}\%                 \\
\hline
\end{tabular}
\caption{Total share of CDNs for Internet services \cite{proxy}}
\label{table:auths-market}
\end{table}

While $76\%$ of Authoritative servers do not use CDNs, among the other $24\%$, Cloudflare is used by $79.8\%$ and Fastly by $6.9\%$, see Table \ref{table:auths-market}. Thus, Cloudflare is the biggest to use \DNSUD{}, but its total share ($19.1\%$) is still relatively small.
}%%%%%%%%%%%%%%%%%%%%%%%%%%%%%%%%%%%%%%%%%%%%%%%%%%%%%%%%%%%%%%%%
%%%%%%%%%%%%%%%%%%%%%%%%%%%%%%%%%%%%%%%%%%%%%%%%%%%%%

%%% Local Variables:
%%% mode: latex
%%% TeX-master: "cwc"
%%% End:
\section{\DNSUD{}-\MakeLowercase{ext} additional material}
%%%%%%%%%%%%%%%%%%%%%%%%%%%%%
\label{sec:ApExt}
\subsection{Adding a new resolver to the extension}
\label{sec:ext-joining-process}
Resolvers that recover from a crash for less than $MaxHistory$ time perform a sequence of UpDNS requests-replies until obtaining all the records in the \UpdateDBext{} up to the current time, see Section \ref{sec:update-process}.
However a new resolver or one that lost its network connection for more than $MaxHistory$ time performs the joining process, as in Section \ref{sec:joining-process}. 
In that process the resolver clears its cache and queries the \UpdateDBext{} to receive only records that have been inserted after it has started the joining process. Thus, in the peculiar case that an authoritative server crashes before a resolver starts the joining process (after being down for more than $MaxHistory$ time), this resolver does not receive the updated information that corresponds to the crashed authoritative server. 

%%%%%%%%%%%%%%%%%%%%%%%%%%%%%
\subsection{Security of the \Extension{}}
\label{sec:ext-security}
%%%%%%%%%%%%%%%%%%%%%%%%%%%%%
The additional threat on \Extension{} on top of the threats on the \DNSUD{} is the cache poisoning attack \cite{cache-injection-study, poison1, poison2, poison3} since the \UpdateDBext{} stores also record updates. To mitigate the attack, each authoritative server owner sends a DNSSEC \cite{rfc4035, dnssec-view, dnssec-view2} signature with the record in the InsertDomain request. Then, upon receiving the UpDNS response by the resolver, it verifies the signature of the record and then inserts the new DNS record to its local cache.
\ignore{%%%%%%%%%%%%%%%%%%%%%%%%%%%%%%%%%%%%%%%%
\begin{enumerate}[leftmargin=0.15in]\setlength\itemsep{0.25em}
    \item Integrity of records stored by \UpdateDB{} - An attacker can cause a DNS cache poisoning attack\cite{poison1, poison2, poison3} by injecting the updated DNS records with incorrect values. The possible damage is high since Internet services may increase their TTLs indefinitely.
    \item DoS/DDoS on the \Extension{} that affects its availability - This becomes more severe since it prevents website owners from updating their DNS records if their authoritative server became inaccessible.
\end{enumerate}
}%%%%%%%%%%%%%%%%%%%%%%%%%%%%%%%%%%
\nsdionly{%%%%%%%%%%%%%%%%%%%%%%%
\subsection{DNS unicast routing support} 
\label{sec:ext-unicast}
\delete{
Unlike the \DNSUD{} service, \Extension{} does not allow Internet services to use unicast DNS services after a record update, since the authoritative server owner can send only one record to the \UpdateDBext, which distributes to all resolvers. Thus, Internet service owners may use anycast DNS services or wait until their authoritative servers become available again.
}
}

\ignore{%%%%%%%%%%%%%%%%%%%%%%%%%%
We suggest the following mitigation to deal with this threat:\newline
Signing DNS records and not only the communication. Authoritative servers should sign them with DNSSEC  to prevent the injection of an incorrect records. Then, each resolver verifies the signature of a record, and only after the verification, it stores the received new IP address for a domain name in its local cache.
}%%%%%%%%%%%%%%%%%%%%%%%%%%%%%%%%%
%%%%%%%%%%%%%%%%%%%%%%%%%%%%%

%% file: client.tex
\section{The client side}
\label{sec:beyond}

%%%%%%%%%%%%%%%%%%%%%%%%%%%%%%%%%%%%%%%%%%%%%%%%%%%%%%%%%%%%%%%%%%%%%%%%%%%%%%%%%%%%%%%%%
%%%%%%%%%%%%%%%%%%%%%%%%%%%%%%%%%%%%%%%%%%%%%%%%%%%%%%%%%%%%%%%%%%%%%%%%%%%%%%%%%%%%%%%%%%%
%%%%%%%%%%%%%%%%%%%%%%%%%%%%%
%%%%%%%%%%\subsection{Should clients communicate directly with the \UpdateDB?}
%%%%%%%%%%\label{sec:traffic-savings}
%\label{sec:}
%%%%%%%%%%%%%%%%%%%%%%%%%%%%%%%
The mechanism described in Section \ref{sec:solution} explains how to promptly push new updates from authoritative servers to resolvers, while maintaining large TTL values.
We discuss here possible ways to also enable clients and stub resolvers to refresh erroneous records in their cache.
Three possible methods may be considered:
\begin{enumerate}%%[leftmargin=0.15in]\setlength\itemsep{0.25em}
    \item
    Enable clients to communicate directly with \UpdateDB{} in \DNSUD{}, thus bypassing the resolvers. 
    This is infeasible since \UpdateDB{} would have to serve  billions of endpoints simultaneously.
    Moreover, the small cache size of clients \cite{cache-size} is not suitable for efficiently working with \DNSUD. 
    \item
    A possible approach is to reduce the traditional TTLs, thus requiring endpoints to frequently get fresh records from the resolvers.
    While this increases the DNS traffic between endpoints and resolvers, i.e., increasing the load on resolvers, it still maintains the \DNSUD{} advantages and reduces the load and traffic on the authoritative servers.
    \item Finally, here we suggest that when a client (e.g., browser) fails to establish a TCP connection with a domain or receives an HTTP error, it may suspect that it has the wrong IP address. 
    In such cases, the browser ignores its cache and issues a new DNS query to its resolver to obtain a fresh record of the website's domain.
    Also we recommend programming the refresh button to delete the corresponding local DNS cache entry and requery that domain automatically.
    Note that this expansion requires updates to billions of endpoints, some of which, like browsers, are frequently and automatically updated, while for others it might necessitate a lengthy adaptation period.
    This, combined with \DNSUD{}, would achieve a fast recovery time from an erroneous update while reducing the futile DNS traffic between all DNS components.
\end{enumerate}

\subsection{Expanding \DNSUD{} to the client side}
\label{sec:beyond-exp}
%%%%%%%%%%%%%%%%%%%%%%%%%%%%%
Here, we elaborate on the 3rd method suggested above.
When a client (e.g., browser) fails to establish a TCP connection with a domain or receives an HTTP error, it may suspect that it has the wrong IP address. 
In such cases, the browser ignores its cache and issues a new DNS query to its resolver to obtain a fresh record of the website's domain.
Additionally, we recommend programming the refresh button to delete the corresponding local DNS cache entry and requery  that domain before refreshing the page.
Note that this expansion requires updates to billions of endpoints, some of which, like browsers, are frequently and automatically updated, while for others it might necessitate a lengthy adaptation period.

\nsdionly{%%%%%%%%%%%%%%%%%%%%%%%%%%%%%%%%%%%%%%%%%%%%%%%%%%%
\begin{figure}[h]
    \centering
    \includegraphics[width=8cm]{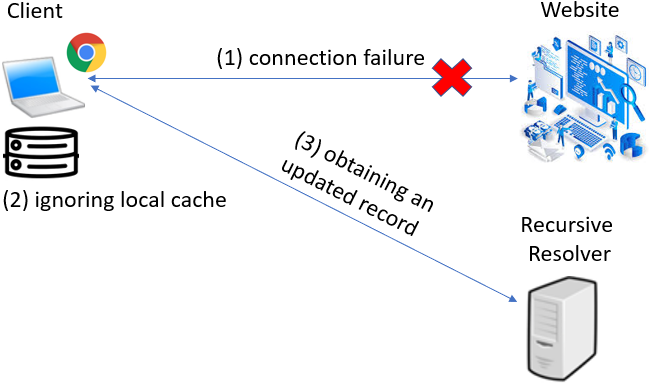}
    \caption{\DNSUD{} expansion overview}
    \label{fig:clients-resolvers}
\end{figure}
}%%%%%%%%%%%%%%%%%%%%%%%%%%%%%%%%%%%%%%%%%%%%%%%%%%%%%%%%%
\ignore{
This, combined with \DNSUD{}, would achieve a fast recovery time for Internet services from an erroneous update while reducing the futile DNS traffic between all DNS components.}

%%%%%%%%%%%%%%%%%%%%%%%%%%%%
\ignore{%%%%%%%%%%%%%%%%%%%%%%%%%%%%%%%%%%%%%%%%%%%%%%%%%%%%%%%%%%%%%%%%
\subsection{Undiscussed implementation issues}
\ariel{Implementing \DNSUD{} raises some issues not discussed here, e.g., what is the exact process in which a website owner registers for \DNSUD{}? Is it paid? etc. All of these essential questions need to be answered by the system implementer.}
}%%%%%%%%%%%%%%%%%%%%%%%%%%%%%%%%%%%%%%%%%%%%%%%%%%%%%%%%%%%%%%%%%%%%%
%%%%%%%%%%%%%%%%%%%%%%%%%%%%

\ignore{%%%%%%%%%%%%%%%%%%%%%%%%%%%%%%%%%%%%
In chapter \ref{sec:analysis} we show that \DNSUD{} is suitable for the majority of the most popular websites. Additionally, we measured the traffic load on the \UpdateDB{} service to show that the proposed method allows a scalable solution for Internet service owners.
Although, \DNSUD{} does not provide a recovery solution for every website. Mainly, the method is not suitable for Internet services that need to change IP addresses frequently or want a very short time (e.g., less than $\Delta$) for total recovery. These services may simply continue with the current DNS resolution process.}%%%%%%%%%%%%%%%%%%%%%%%%%%%%%%%%%%
%%% Local Variables:
%%% mode: latex
%%% TeX-master: "cwc"
%%% End:

%% file: discussion.tex
\section{Relevant CDN market trend}
\begin{figure}[h]
    \centering
    \includegraphics[width=8cm]{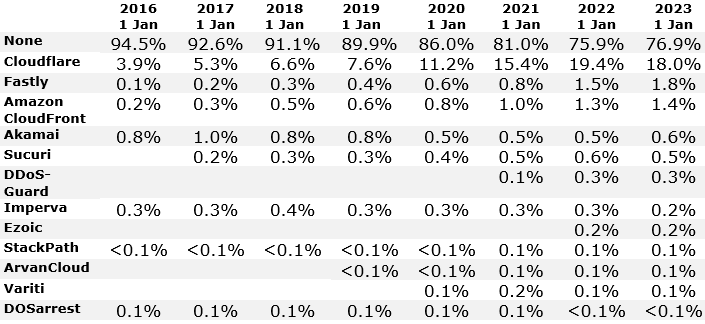}
    \caption{Market share trends for CDN services from W$^3$Techs \cite{cdn-market}}
    \label{fig:marketshare}
\end{figure}

%Discussion - why now?}
\label{sec:discussion}
\ignore{%%%%%%%%%%%%%%%%%
Recent trends in the CDN market make the \DNSUD{} more relevant currently. 
The problem presented in this paper is not new; it has been troubling for many years, and yet, as far as we know, there is no solution for it "in the wild."
In recent years, a CDN hosting trend has emerged that makes \DNSUD{} applicable:

Major global CDN providers (like Cloudflare and Fastly) offer more stable IPs for domains they service, relying on fewer edge data-centers to route users to the optimal data center internally rather than reflecting those changes in the resolved IP. 
According to Figure \ref{fig:marketshare}, the market share of CDNs is rising, leading to an increasing number of domains becoming \emph{stable}. 
Additionally, as we observed, over $80\%$ of domains are \emph{stable}, meaning that even the majority of domains not serviced by CDN services are \emph{stable}.
Therefore, the resulting increase in \emph{stable} domains makes \DNSUD{} an attractive mechanism currently.
}%%%%%%%%%%%%%%%%%%%%%%%
Two recent trends elevate the portion of \emph{stable} domains thereby enhancing the relevance of \DNSUD{}.
First, major global CDN providers, such as Cloudflare and Fastly, offer more \emph{stable} IPs for the domains they service, since they rely on a fewer number of edge data-centers to route users optimally, instead of using a large number of reverse proxies and reflecting the routing decisions on the resolved IP address (as Akamai).
Secondly, as depicted in Figure \ref{fig:marketshare}, the market share of CDNs is on the rise.
Thus. the percentage of \emph{stable} domains increased to over $80\%$, as seen in Section \ref{sec:freq-updates}.
\ignore{%%%%%%%%%%%%%%%%%
Consequently, the resulting rise in \emph{stable} domains currently makes \DNSUD{} an increasingly attractive mechanism.
we believe it is the right time to explore a solution that is scalable, feasible, efficient, secure, backward-compatible, and requires minimal changes from the standard DNS system.
}%%%%%%%%%%%%%%%%%%%%%%%

%% file: conclusion.tex
\section{Conclusions}
\label{sec:conclusion}
We introduce \DNSUD, a new novel DNS service that enhances the DNS system high-availability by facilitating secure, real-time updates of cached domain records in resolvers, even before the associated TTL has expired.
We examined all implications and considerations of \DNSUD{} that we could think of, arguing the advantages, feasibility, backward comparability and gradual deployment properties of \DNSUD{}.
\DNSUD{} not only enables swift corrections of mistaken domain updates but also promotes longer TTL values, thus reducing futile traffic from both resolver and authoritative server load. 
After \DNSUD{} would be adopted, it is plausible to reevaluate resolver cache sizes, policies, and consider transitioning authoritative servers to scalable services.

The traditional DNS is a "Pull" system where clients and resolvers actively fetch data from authoritative servers. \DNSUD{} introduces a "Push" element to this framework, overcoming some inherent limitations of the conventional "Pull" method.

\nsdionly{%%%%%%%%%%%%%%%%%%%%%%%%%%%%%%%%%%%%
Then, we analyze the feasibility and relevance of the \DNSUD{} service and show that it is relevant to more than $80\%$ of domains and could effectively serve more than $391,000,000$ different domains.
Note that \DNSUD{} is compatible with unicast DNS services \cite{unicast}, meaning it can provide different IP addresses to different resolvers.
}%%%%%%%%%%%%%%%%%%%%%%%%%%%%
\nsdionly{
In summary, \DNSUD{} is relevant to solving the following:
\begin{description}[leftmargin=0.12in]\setlength\itemsep{0em}
    \item [DNS record has been updated:] \DNSUD{} ensures quick recovery time independently of the domain's TTL.
    \item [Authoritative server becomes unavailable:] Clients get the correct IP address for the requested domain, as long as no record update is required due to significantly increasing the TTL associated with the domain.
    \item [After authoritative server becomes unavailable its domain] 
    \textbf{name record has been changed:} Internet service owners can update DNS records through the \UpdateDBext{}.
\end{description}
}
\ignore{Different papers \cite{pushing-dns,eliminating-dns,p2p-dns} suggest different methods with significant DNS protocol or IP system changes. Compared to them,...}%%%%%%%%%%%%%%%%%%%%%%%%%%%%%%%%%%%%%%%%%%%%%%%%%%%%%%%%%%%%%%%%% 
\nsdionly{%%%%%%%%%%%%%%%%%%%%%%%%%%%%%%%%%%%%%%%%%%%%%%%%%%%%%%%%%
We created a backward-compatible service that changes the form of communication between different DNS components as little as possible. Finally, in Section \ref{sec:beyond} we suggest a method that allows clients to receive DNS record updates before the associated TTLs have expired and reduce traffic between clients and recursive resolvers.
}%%%%%%%%%%%%%%%%%%%%%%%%%%%%%%%%%%%%%%%%%%%%%%%%%%%%%%%%%%%%%%

\ignore{%%%%%%%%%%%
\DNSUD{} may be extended to provide remediation even if all authoritative servers for a domain are unavailable while the domain IP address is updating.
While this extension overcomes a rather rate faulty event, it is more complicated than \DNSUD{} and raises new security concerns.
It necessitates storing also the full record for each domain in the \UpdateDB.
Due to space limitations, we relegate discussing this extension, to the full version.

We could extend \DNSUD{} to provide remediation even if all authoritative servers for a domain are unavailable during the domain update. 
This requires farther changes (storing also the full record for each domain in the \UpdateDB)
to the basic version of \DNSUD. Due to space limitations we relegate discussing this extension, to the full version. 
}%%%%%%%%%%%

\nsdionly{
\DNSUD{} is sort of a push method in a pull scheme.
Authoritative servers push domain names to the \UpdateDB{} and in response resolvers pull both the \UpdateDB{} and the authoritative servers.
The pull method has three advantages:
First, it allows easy addition of new resolvers to the \DNSUD{} system.
Secondly, each resolver may determine its method parameters, i.e., $\Delta$, independently.
Finally, it allows the \UpdateDB{} to be stateless - it does not need to maintain a list of resolvers, their last timestamps, etc.

\DNSUD{} can be extended to provide remediation even when all authoritative servers for a domain are unavailable during the domain IP address update. 
However, this extension introduces additional complexities compared to \DNSUD{} and raises new security concerns. 
It necessitates storing the complete record for each domain in the UpdateDB. 
Due to space limitations, a detailed discussion of this extension is deferred to the full version.
}

\ignore{%%%%%%%%%%%%%%%%%%%%%%%%%%%%%%%%%}
\DNSUD{} is sort of a pull method (the resolvers periodically query the \UpdateDB{}) initiated by a push (the authoritative servers insert updated domain name records).
The choice of a pull method has three advantages:
First, it allows easy addition of new resolvers that use \DNSUD{} since a new resolver just needs to implement \DNSUD{}.
Secondly, each resolver can define its method parameters, i.e., $\Delta$.
Finally, it allows the \UpdateDB{} to be stateless - it does not have to maintain a list of resolvers, their last timestamps, etc.
}%%%%%%%%%%%%%%%%%%%%%%%%%%%%%%%%%%%%%%%%

%%% Local Variables:
%%% mode: latex
%%% TeX-master: "cwc"
%%% End:

%% file: acknowledgements.tex
\section*{acknowledgements}
%%We would like to thank SecurityTrails \cite{securitytrails} for the historical DNS information of IP changes.
This research is supported by the Blavatnik Computer Science Research Fund.
We would like to thank Anat Bremler-Barr for very helpful feedback throughout the research and on earlier versions. 
We also thank SecurityTrails \cite{securitytrails} for the historical DNS information of IP changes.